\documentclass[prx,twocolumn,floatfix,superscriptaddress]{revtex4-1}

\usepackage{amssymb,amsmath,amstext}
\usepackage{graphicx}
\usepackage{epstopdf}
\usepackage{color}
\usepackage{bm}
\usepackage{appendix}
\usepackage[T1]{fontenc}
\usepackage{bbold}
\usepackage{bbm}
\usepackage{latexsym}
\usepackage{ulem}

\usepackage{physics}
\usepackage{braket}
\usepackage{dsfont}
\usepackage{float}
\makeatletter
\let\newfloat\newfloat@ltx
\makeatother
\usepackage{algorithm}
\usepackage{algpseudocode}

\usepackage{xcolor}
\usepackage{adjustbox}
\usepackage{tikz}
\usetikzlibrary{quantikz}

\usepackage{subcaption}

\usepackage[colorlinks=true,citecolor=blue,linkcolor=magenta]{hyperref}
 
\usepackage{cleveref}

\newcommand{\identity}{\mathds{1}} 
\newcommand{\imag}{\mathrm{i}}

\newcommand{\half}{\frac{1}{2}}
\newcommand{\argHalf}[1]{\frac{#1}{2}}

\newcommand{\clascirc}{\ensuremath{M(\tau)}}

\newcommand{\cliffordgroup}{\ensuremath{{\mathcal{C}}}}

\newcommand{\numqubits}{\ensuremath{{N_{\text{q}}}}}
\newcommand{\numel}{\ensuremath{{N_{\text{el}}}}}
\newcommand{\numclusters}{\ensuremath{{N_{\text{cl}}}}}

\newcommand{\balpha}{{\boldsymbol{\alpha}}}

\newcommand{\review}[2]{#2}

\renewcommand{\hypertarget}[2]{\label{#1}}
\renewcommand{\hyperlink}[2]{#2~in~\cref{#1}}

\begin{document}

\title{Partitioning Quantum Chemistry Simulations with Clifford Circuits}

\author{Philipp Schleich}
\affiliation{Department of Computer Science, University of Toronto, Canada.}
\affiliation{Vector Institute for Artificial Intelligence, Toronto, Canada.}
\affiliation{Theoretical Division, Los Alamos National Laboratory, Los Alamos, NM, USA}

\author{Joseph Boen}
\affiliation{Department of Applied Mathematics \& Statistics, Johns Hopkins University, Baltimore, MD, USA}
\affiliation{Theoretical Division, Los Alamos National Laboratory, Los Alamos, NM, USA}

\author{Lukasz Cincio}
\affiliation{Theoretical Division, Los Alamos National Laboratory, Los Alamos, NM, USA}

\author{Abhinav Anand}
\affiliation{Chemical Physics Theory Group, Department of Chemistry, University of Toronto, Canada.}
\affiliation{Duke Quantum Center, Duke University, Durham, NC 27701, USA}
\affiliation{Department of Electrical and Computer Engineering, Duke University, Durham, NC 27708, USA}

\author{Jakob S. Kottmann}
\affiliation{Department of Computer Science, University of Toronto, Canada.}
\affiliation{Chemical Physics Theory Group, Department of Chemistry, University of Toronto, Canada.}


\author{{Sergei Tretiak}}
\affiliation{Theoretical Division, Los Alamos National Laboratory, Los Alamos, NM, USA}
\affiliation{
Center for Integrated Nanotechnologies, Los Alamos National Laboratory, Los Alamos, NM, USA}

\author{{Pavel A. Dub}}
\affiliation{Chemistry Division, Los Alamos National Laboratory, Los Alamos, NM, USA}

\author{Al\'{a}n Aspuru-Guzik}
\email[]{aspuru@utoronto.ca}
\affiliation{Chemical Physics Theory Group, Department of Chemistry, University of Toronto, Canada.}
\affiliation{Department of Computer Science, University of Toronto, Canada.}
\affiliation{Department of Chemical Engineering and Applied Chemistry,  University of Toronto, Canada.}
\affiliation{Department of Materials Science and Engineering, University of Toronto, Canada.}
\affiliation{Vector Institute for Artificial Intelligence, Toronto, Canada.}
\affiliation{Canadian  Institute  for  Advanced  Research  (CIFAR)  Lebovic  Fellow,  Toronto,  Canada.}

\begin{abstract}
Current quantum computing hardware is restricted by the availability of only few, noisy qubits which limits the investigation of larger, more complex molecules in quantum chemistry calculations on quantum computers in the near-term.
In this work, we investigate the limits of their classical and near-classical treatment while staying within the framework of quantum circuits and the variational quantum eigensolver.
To this end, we consider naive and physically motivated, classically efficient product ansatz for the parametrized wavefunction adapting the separable pair ansatz form.  We combine it with post-treatment to account for interactions between subsystems originating from this ansatz. 
The classical treatment is given by another quantum circuit that has support between the enforced subsystems and is folded into the Hamiltonian. To avoid an exponential increase in the number of Hamiltonian terms, the entangling operations are constructed from purely Clifford or near-Clifford circuits. While Clifford circuits can be simulated efficiently classically, they are not universal. In order to account for missing expressibility, near-Clifford circuits with only few, selected non-Clifford gates are employed. The exact circuit structure to achieve this objective is molecule-dependent and is constructed using simulated annealing and genetic algorithms. We demonstrate our approach on a set of molecules of interest and 
investigate the extent of our methodology's reach.
\end{abstract}
\maketitle

\section{Introduction} 
An important goal of computational chemistry is to understand the behavior of strongly correlated electronic systems, as they are key to many applications in materials science. Examples hereof are the development of more efficient catalysts, drugs or high-temperature superconductors.
Since the computational complexity of many existing classical, exact numerical methods scales exponentially with system size owing to solving the Schr\"odinger equation, there has been increased interest in developing quantum algorithms for this task in recent years~\cite{aspuru2005simulated,cao2019quantum,mcardle2020quantum}. One of the most popular approaches for finding the ground state~\cite{peruzzo2014variational} and several low-lying excited states~\cite{higgott2019variational,nakanishi2019subspace,mcclean2017hybrid,asthana2023quantum} is the Variational Quantum Eigensolver (VQE)~\cite{cerezo2021variational,tilly2021variational,bharti2022noisy}.

The VQE is a hybrid quantum-classical algorithm that utilizes the variational principle to find variational approximations to the eigenstates and associated eigenvalues of a molecular Hamiltonian.  
Given a molecular Hamiltonian encoded as a linear combination of tensor products of Pauli operators, a parameterized quantum circuit (PQC) is used to prepare a trial wavefunction (ansatz) and compute its energy. 
By optimizing the parameters of the quantum circuit, approximations to the ground and also excited states and their energies can be obtained.
The VQE has been successfully applied to simulations of small molecules on quantum hardware \cite{kandala2017hardware}. 
One strategy to find trial wavefunctions is given by chemically-inspired ansatze, such as the popular family of Unitary Coupled Cluster (UCC) ansatze~\cite{peruzzo2014variational, anand2021quantum}. 
UCC utilizes domain knowledge from quantum chemistry to prepare trial wavefunctions by exciting electrons from a reference state, which is usually the Hartree-Fock state. 
In contrast, hardware-efficient ansatze~\cite{kandala2017hardware, ollitrault2020hardware} construct circuits from a limited set of gates that exploit the characteristics of the underlying device, such as connectivity or coherence time. Instead of directly encoding chemical knowledge into the ansatz, hardware-efficient ansatze rather try to be maximally expressible given hardware constraints. 
Despite continuous development of quantum hardware, current devices are prone to significant errors and can only support a limited number of operations on a few qubits, which limits the complexity of the quantum circuits used to generate trial wavefunctions. 
Thus, the development of techniques for designing shallow and resource-efficient ansatze is not only academically significant, but also of considerable practical importance.
High gate errors and low coherence times of devices that are available in the current and near future -- so called noisy intermediate-scale quantum computers (NISQ)~\cite{preskill2018quantum} -- pose a critical limit to the circuit depth. As a result, many techniques have been developed to find suitable shallow circuits, see refs.~\citenum{bharti2022noisy, cerezo2021variational, anand2021quantum} for a representative overview. 

\begin{figure}[h!]
    \centering
    \includegraphics[width=0.975\linewidth]{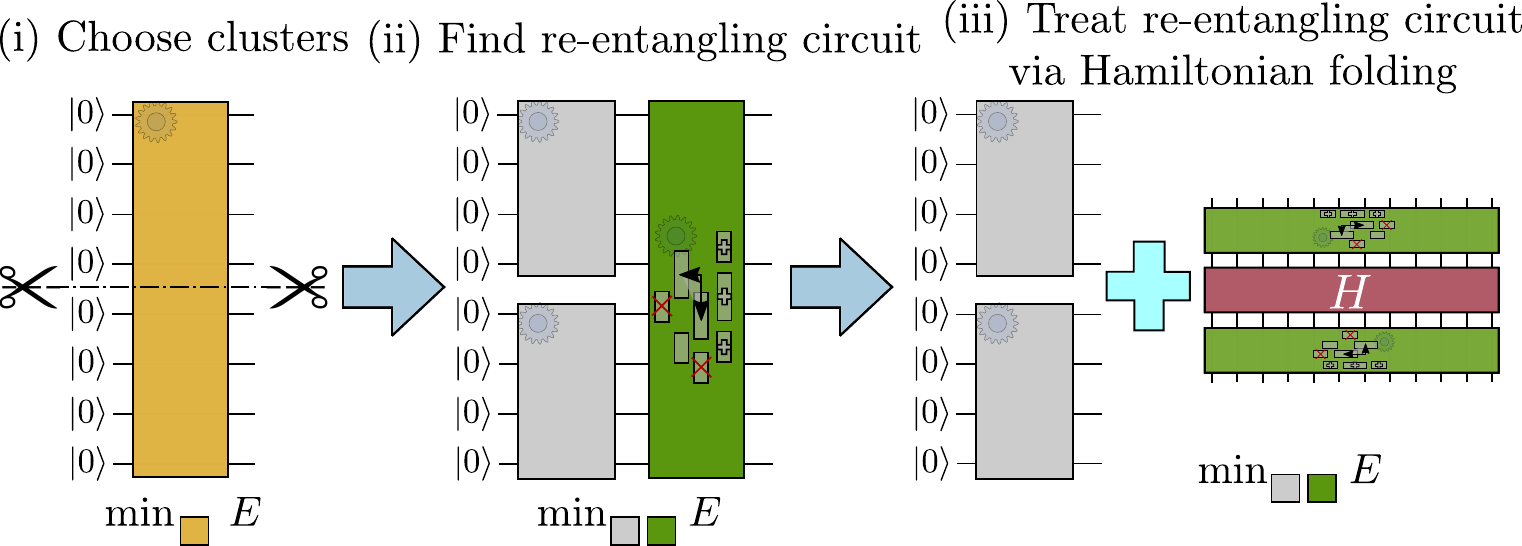} 
    \caption{
    Overview of our proposed method of dividing large system into smaller pieces. The ansatz consists of two parts: (i) VQE-like circuits denoted by gray boxes and with continuous, variational parameters (ii) near-Clifford, classically simulable circuit (green box) whose architecture is to be found by the optimizer as well. Near-Clifford circuits are evaluated by modifying (or dressing) the Hamiltonian $H$. }
    \label{fig:cartoony-overview}
\end{figure}

In this work, we attempt to clusterize circuits to facilitate the usage of existing NISQ devices to run larger experiments.
There has been recent work in this direction, as in refs.~\citenum{zhang2021variational,eddins2022doubling}. 
Both studies attempt to discover clusters or reduce the number of qubits per circuit through the utilization of correlation metrics. 
While ref.~\citenum{zhang2021variational} makes use of mutual information to clusterize the circuit, ref.~\citenum{eddins2022doubling} uses entanglement forging to divide the circuit into two pieces that are executed separately.  In our method, we pre-define the clusters and generally ``cut'' the circuit into two equally large pieces. 
We consider two options in choosing the clusters. For a physically motivated choice,  we use the separated-pair-ansatz from ref.~\citenum{kottmann2022optimized}, which is classically tractable and provides a natural clustering into electron-pairs, made up of pair-natural orbitals~\cite{kottmann2021reducing,kottmann2020direct}. Beyond that, we also consider arbitrary clusters. Here, we expect that the sub-optimal choice of clusters can be accounted for by a Clifford circuit since permutation of qubits can be realized as a sequence of Clifford gates.
Then, to re-gain lost correlation by assuming the product state, we search for a re-entangling circuit.%
In order to enable that the quantum computation can be carried out on hardware with a lower number of qubits, we use this circuit to similarity-transform the system Hamiltonian. In the literature\cite{ryabinkin2020iterative}, this transform is often called ``folding'' or ``dressing'' the Hamiltonian. This procedure then is similar to the virtual Heisenberg circuits from ref.~\citenum{shang2021schr} or ref.~\citenum{zhang2021variational}. The latter optimizes, beyond the quantum and the virtual circuit, a classical neural network to enable error mitigation.%
The circuit that modifies the Hamiltonian is trained together with VQE circuits, minimizing the overall energy. This means that the (near-)Clifford circuit modifies the Hamiltonian so that its ground state is closer to a product state.
This procedure is briefly outlined in \cref{fig:cartoony-overview}. We emphasize here, that both the cluster circuits as well as the virtual circuits are to be optimized. This means that for each modification of the virtual circuit, the parameters in the cluster circuits will be adjusted. 

For the re-entangling circuits, we will first focus on using Clifford circuits only. The latter describe sequences of unitaries composed of elements of the Clifford group. Thus, they leave the number of Pauli strings when transforming a Hamiltonian in the Pauli basis invariant. As a consequence, the measurement complexity does not increase as opposed to folding general gates into the Hamiltonian. This will be extended by near-Clifford circuits, which introduce a small number of non-Clifford operations into these gate sequences so that the expressibility is increased but the operator complexity does not increase vastly. It is known that Clifford circuits are efficiently simulatable classically but do not form a universal set of quantum gates \cite{gottesman1998theory, gottesman1998heisenberg, bravyi2005universal}. 
There have been recent efforts for the usage of Clifford circuits on early-stage quantum computers, such as the development of initialization strategies to find better initial circuits and guesses for gate parameters based on Clifford gates only \cite{mitarai2020quadratic, cheng2022clifford}. Other areas of application include classically efficient compression of quantum circuits~\cite{anand2022quantum} and a standalone ansatz for variational quantum algorithms~\cite{ravi2022cafqa}. Such a parametrization however is not able to reproduce the results obtained with universal gate sets due to limited expressibility of the Clifford group. 
In this study, we are exploring the limit to which classical post-processing (mostly relying on Clifford gates) can reduce the resource requirements in quantum chemistry calculations. 
Since Clifford gates form a discrete set, the optimization requires specialized techniques, as described\hyperlink{subsec:optimizing-cliffords}{}.

We start this work by outlining our proposed methodology in~\cref{sec:methodology}, including a motivation for  the choice of clusters, looking at how to fold important circuit identities and the way Clifford and near-Clifford circuits are constructed. Then, we show \hyperlink{sec:results}{results} for simulating the ground-states of several molecules (H$_2$, N$_2$ and BeH$_2$)  before \hyperlink{sec:conc-and-outlook}{concluding our work}.

\section{Methodology}\label{sec:methodology}
\subsection{Outline}
In a VQE calculation, one aims to find an approximation $E$ to the ground-state energy $E_0$
\begin{equation}\label{eq:minenergy}
    E_0 \le E = \min_{\psi} \braket{ \psi | H | \psi }.
\end{equation}
The state $\ket{\psi}$ is usually prepared and optimized using a parametrized unitary operation $\ket{\psi(\theta)} = U(\theta) \ket{0}$, where $U(\theta)$ acts on $\numqubits$ qubits.

\begin{figure*}
    \centering
    \includegraphics[width=\textwidth]{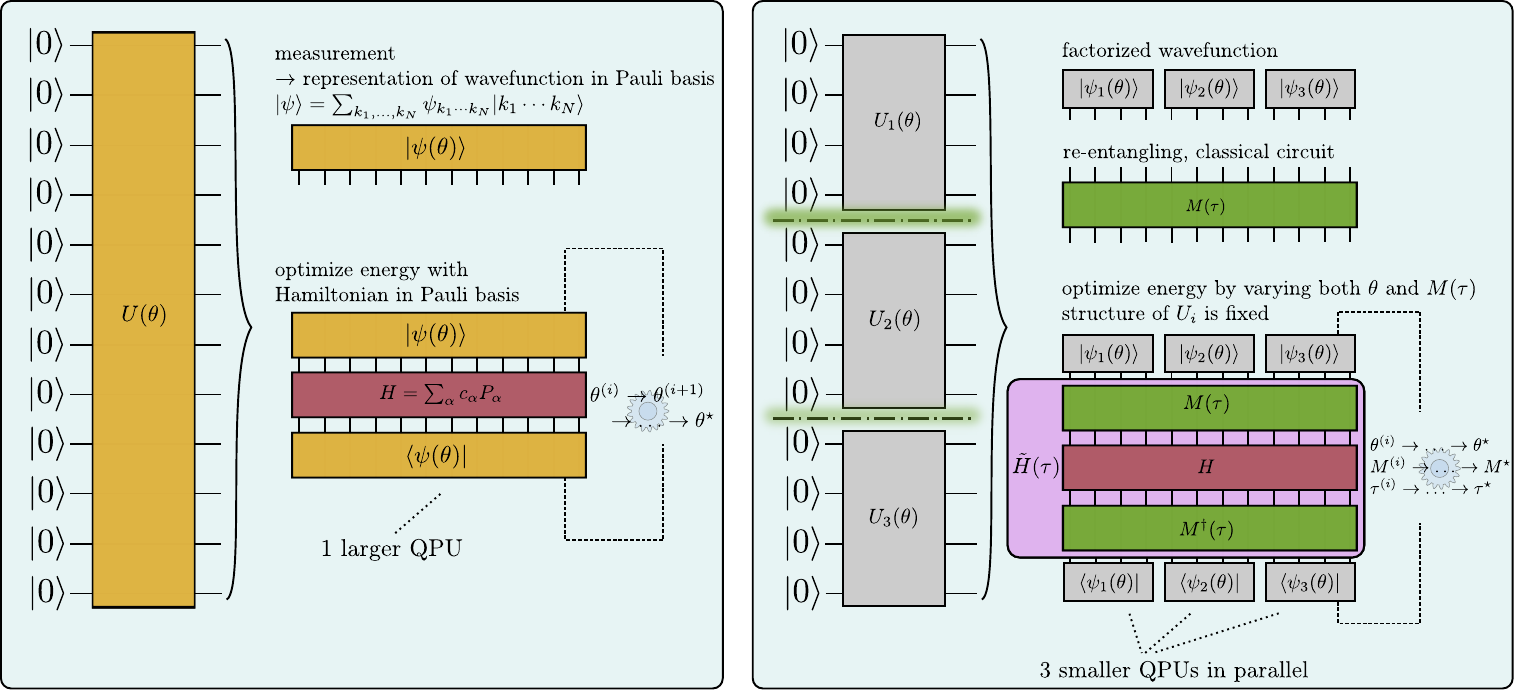}
    \caption{Comparison of standard VQE procedure (left) and our approach (right), depicted using an example of three clusters. }
    \label{fig:main-overview}
\end{figure*}

In this work, we prepare the wavefunction $\ket{\psi(\theta)}$ through unitaries $U_j(\theta_j)$ that only have local support on qubit-cluster $j$ \review{(cluster-local)}{} followed by a global operation $\clascirc$ that is built using a (near-)Clifford circuit, which  
add some additional variational parameters~$\tau$:
\begin{align}\label{eq:classical-cluster-ansatz}
    \ket{\psi(\theta, \tau)} &= \clascirc \prod_{j=1}^{\numclusters} \underbrace{U_j(\theta_j) \ket{0}}_{=: \ket{\psi_j(\theta_j)}}.
\end{align}
Here, $\clascirc$ acts on the full space over $\numqubits$ qubits, while $U_j$ act only on the respective clusters $j=1,\ldots,\numclusters$ with $\numqubits^{[j]}$ qubits.  
This way, we have a set of $\numclusters$ factorized states that can be evaluated independently and are then re-combined with $\clascirc$. This circuit $\clascirc$ will be folded into the Hamiltonian. Our method can be summarized as a Schr\"odinger evolution on a subsystems with a virtual Heisenberg circuit that alters the Hamiltonian so that its ground-state resembles a product state over the subsystems. We provide an outline of the procedure in comparison with the standard, Schr\"odinger-picture-like VQE, in \cref{fig:main-overview}. 

In what follows, we will first discuss how one can \hyperlink{subsec:clustering}{choose the subsystems}. 
Then, as the main contribution of this work, we will outline \review{}{\hyperlink{subsec:recovering-virtual-heisenberg}{our procedure}} to find the virtual circuit  $\clascirc$ \review{in Section~\ref{subsec:recovering-virtual-heisenberg}}{} so that one regains maximum correlation energy.

\subsection{Clustering into subsystems}\hypertarget{subsec:clustering}{}
\review{We choose subsystems either in a naive, uninformed, or physically motivated manner.}{We choose subsystems either in a naive, unformed manner or guided by physical intuition.}

In  the naive approach, we set clusters by simply cutting the circuit in $\numclusters$ roughly equally large clusters without considering any additional information. This allows us to test the capabilities of the re-entangling circuit $M(\tau)$ to correct \review{}{for} shortcomings of the naive clustering compared to the physically motivated approach and experiment with more arbitrary clusterings - for example, within electron pairs. 
In particular, we will see that the inclusion of permutations~\cite{tkachenko2021correlation} into the ansatz allows the approach to retrieve a more optimal clustering.

For the physically motivated splitting, we use the separated-pair ansatz (SPA)~\cite{kottmann2022optimized} with fixed pair-natural orbitals -- i.e., we are not considering orbital-optimization as in ref.~\citenum{kottmann2022molecular}.
The SPA ansatz admits a natural decomposition into factorized clusters made up by $\numel/2$ electron pairs, where the full wavefunction reads (cp. Eq.~(15) in ref.~\citenum{kottmann2022optimized})
\begin{equation}
    \ket{\psi_\textrm{SPA}} = \prod_{j=1}^{\numel/2} U_j(\theta_j) U_\textrm{HF} \ket{00\cdots 0} \ ,
\end{equation}
with pair-specific UCC-circuits $U_j$, number of electrons $\numel$ and a circuit to prepare the Hartree-Fock state $U_{\text{HF}}$. 

The reasoning behind our method is that we, on one hand, would like to push the boundaries of the classically efficient SPA-ansatz\review{, but also, on the other hand,}{. On the other hand, we also} expect a self-assembly of clusters throughout our technique. The latter can be understood in the sense of the so-called permVQE~\cite{tkachenko2021correlation}, where qubits (or conversely, orbitals) are permuted in order to maximize mutual information leading to a reduction in circuit depth. Since permutations are Clifford operations, including them in the allowed set of operators for $\clascirc$ does not increase complexity of the Hamiltonian. This means that the sub-optimal choice of clusters here is correctable by the virtual circuit $\clascirc$.

As we will discuss \review{ in Section~\ref{subsec:recovering-virtual-heisenberg}}{\hyperlink{subsec:recovering-virtual-heisenberg}{later}}, any modification of the Hamiltonian beyond Clifford gates may increase its complexity in form of the number of independent terms in the Pauli basis significantly.
While it is in general beneficial to use measurement reduction techniques to lighten the measurement load, it can particularly help here when non-Clifford gates in $\clascirc$ increase the measurement complexity.
These procedures typically effectuate a grouping of the Hamiltonian terms by various means so that multiple terms can be measured simultaneously \cite{izmaylov2019unitary,yen2020measuring,verteletskyi2020measurement,bonet2020nearly,yen2021cartan,huggins2021efficient}. 
At this point, we note that most of the techniques find a unitary operation that is applied directly before measurement so that the grouping of terms can be achieved in a desired basis. This is the case in particular for the methodologies outlined in refs.~\citenum{yen2021cartan, huggins2021efficient} that suppress the cost to at most $\mathcal{O}(N^2)$ measurements. Unfortunately, this unitary gate needs to be evaluated on the quantum hardware with support beyond clusters for the method to be effective, which is impractical when dividing the circuit \review{in}{into} clusters with the objective to run each cluster on hardware independently. 
However, one can still make advantage of graph clique cover methods such as ref.~\citenum{verteletskyi2020measurement} and especially the approach proposed in ref.~\citenum{bonet2020nearly}, which can also compress the measurements into a number of partitions that is at best $\order{N^2}$.

\subsection{Recovering correlation by folding a Heisenberg-circuit }\hypertarget{subsec:recovering-virtual-heisenberg}{}

Here, we describe our approach to construct the re-entangling circuit $\clascirc$ as in \cref{eq:classical-cluster-ansatz}, carried out as a virtual Heisenberg circuit in the spirit of refs.~\citenum{zhang2021variational, shang2021schr, zhangB2021variational}. We first start by taking a closer look at folding  typical circuit building blocks, not necessarily Clifford, in a generator formalism before introducing a procedure to build $\clascirc$.
Hereafter, we use the expression ``to fold'' a unitary circuit $U$ into a Hermitian operator $H$ to indicate a mapping according to a similarity transformation $H\overset{U}{\to} U^\dagger H U$. 
We will be using either purely-Clifford or near-Clifford circuits, which contain a small number of non-Clifford, parametrized gates to increase expressibility~\cite{sim2019expressibility} of the circuit.

\subsubsection{Folding of important circuit identities}\hypertarget{subsubsec:folding}{}
We aim to approximate the  ground-state energy of an electronic system using an ansatz specified in \cref{eq:classical-cluster-ansatz} with a virtual circuit $\clascirc$, given a clustering ansatz $\prod_j U_j(\theta)$.
The resulting cost function, as in \cref{eq:minenergy}, then takes the form
\begin{equation}
    \braket{ 0 | (\prod_{j=\numclusters}^{1} U_j^\dagger(\theta)) \underbrace{\clascirc^\dagger H \clascirc}_{=:\Tilde{H}(\tau)}  (\prod_{j=1}^{\numclusters} U_j(\theta)) | 0  }.
\end{equation}
We consider virtual circuits that are built by a sequence of individual unitary gates $\clascirc = \prod_{l=1}^{N_M} V_l (\tau_l)$, where individual operations can be expressed as 
\begin{equation}
    V_l(\tau_l) = \exp\left(-\imag \argHalf{\tau_l} G_l\right)
\end{equation}
for Hermitian generators $G_l=G_l^\dagger$. This allows us to implement the mapping $H \overset{\clascirc}{\longrightarrow} \Tilde{H}(\tau)$ by sequentially folding individual operations $V_l(\tau_l)$.

Since folding operations to find  $V_l^\dagger H V_l$ are an important part of our method (cp. to \cref{fig:main-overview}), we first look at efficient ways to find analytical implementations, using different classes of unitary gates. 
This encompasses recent work~\cite{kottmann2021feasible,izmaylov2021analytic,wierichs2022general,kyriienko2021generalized} related to analytical gradients for parametrized quantum circuits using the parameter-shift rule \cite{schuld2019evaluating}, where in this work, we resort to refs.~\citenum{izmaylov2021analytic,kottmann2021feasible} in particular. 
We first investigate rotation-like gates, whose generators are involutions, and gates that are generated by projections. Both have in common that their generators have exactly two eigenvalues, while the latter can be reformulated as the former by modifications so that their spectrum is symmetric around zero. Beyond that, we discuss excitation-like gates, whose spectrum differs from involutions  by an additional zero-eigenvalue, and discuss a strategy for general gates that is inspired by ref.~\citenum{izmaylov2021analytic}.  

\paragraph{Rotation-like gates.}
For rotation-like gates such as $R_X,R_Y,R_Z$, the generators are involutions $G^2=\identity$. Hence  we can write $V(\tau) = \cos(\argHalf{\tau})\identity -\imag \sin(\argHalf{\tau})G$. This gives us the identity 
\begin{align}\label{eq:rotlike-folding}
    V^\dagger(\tau)HV(\tau) = \cos^2\left(\argHalf{\tau}\right)H &+ \sin^2\left(\argHalf{\tau}\right)GHG \notag\\  &-  \imag \cos(\argHalf{\tau})\sin(\argHalf{\tau})[H,G]. 
\end{align}

\paragraph{Projector-generated gates.}
Another important class of gates is generated by projection operators, i.e. $G^2=G$.
We can motivate this class in the following way: 
If one tries to generate an $X$-gate by using the rotation $R_X(\pi)=\exp(-\imag \argHalf{\pi} X)$, one obtains $-\imag X $. A phase-true $X$ is generated by the projection $G=\half(\identity-X)$ and $\tau=2\pi$, i.e. $X = \exp(-\imag \argHalf{\pi} (\identity - X))$. 
One can thus generally relate this to the class of rotation-like gates from the above section with self-inverse generators $\tilde{G}$ and associated angles $\tilde{\tau}$ by setting $\tau=2\tilde{\tau}$ and $G=\half(\identity-\tilde{G})$:
\begin{equation}
  V(\tau) =\exp(-\imag\argHalf{\tau}G) \equiv \exp(-\imag\argHalf{\tilde{\tau}}\tilde{G}). 
\end{equation}
This means that projector-generated gates are equivalent up to a $U(1)$ transformation to rotation-like gates since $\tilde{G}^2=\identity$. We still deem it worthy to mention them as a separate class to emphasize that a phase-true implementation of certain gates requires a projection as generator, which is of importance for controlled versions of these gates. 
Hence upon having done this substitution, the folding can be performed just as in \cref{eq:rotlike-folding}.

\paragraph{Excitation-like gates.}
For quantum chemistry applications, circuit elements corresponding to fermionic excitations are of particular interest as they conserve particle-number symmetry. Beyond that, qubit excitations \cite{ryabinkin2018qubit,ryabinkin2020iterative,xia2020qubit,anand2021quantum} or (multi)-controlled rotations belong to this class. These gates belong to a class of operations that are generated by $G=\mathds{P}_{+} -  \mathds{P}_{-} + \mathds{P}_0$, projecting on eigenstates with symmetric eigenvalues $\pm \lambda$ and $0$. One may express the generated unitary operator as \cite{kottmann2021feasible}
\begin{equation}
   V(\tau) = \cos(\argHalf{\tau})\identity - \imag \sin(\argHalf{\tau}) G + \Big(1-\cos(\argHalf{\tau})\Big)\mathds{P}_0. 
\end{equation}
The associated folded Hamiltonian is then 
\begin{align}
    V^\dagger(\tau) H V(\tau)  =& \cos^2\left(\argHalf{\tau}\right) H
    +\sin^2\left(\argHalf{\tau}\right)
 {G}H{G} \\ 
 &+  \left((1-\cos\left(\argHalf{\tau}\right)\right)^2 \mathds P_0 H \mathds P_0 \nonumber\\
    &- \imag \cos\left(\argHalf{\tau}\right) \sin\left(\argHalf{\tau}\right) \comm{H}{{G}} \\
    &+ \cos\left(\argHalf{\tau}\right)  \left(1-\cos\left(\argHalf{\tau}\right)\right) \left\{H,\mathds P_0\right\} \nonumber\\
    &+ \imag \sin\left(\argHalf{\tau}\right)  \left(1-\cos\left(\argHalf{\tau}\right)\right) \left({G}H \mathds P_0 - \mathds P_0 H {G}\right).
\end{align}

\paragraph{General gates.}
To obtain an exact expression for general gates, one can use the spectral decomposition of Hermitian generators $G = \sum_k \lambda_k \mathds{P}_k$, with $\lambda_k$ eigenvalues and  $\mathds{P}_k$ projectors on the respective eigenspaces and then proceed similarly as for excitation-like gates, e.g. following the procedure outlined in section B.2. of ref.~\citenum{izmaylov2021analytic}. If the dimensionality of the generator prohibits a numerical eigendecomposition, the gate may be decomposed into lower-dimensional building blocks. 

\paragraph{Hamiltonian-dependent identities.}
Whenever the transformed Hamiltonian has specific properties they might be exploited for similarity transformations with specific gates. 
A prominent example relevant in the context of this work are fermionic single-excitations with $G_{pq} = -i\kappa_{pq} a^\dagger_p a_q$ and anti-Hermitian matrix 
$\kappa$ in combination with Hamiltonians consisting of fermionic annihilation (creation) operators $a_p$ ($a^\dagger_p)$ only. Here, the operators are transformed as
$\tilde{a}_p^\dagger \equiv e^{iG}a_p e^{-iG} = \sum_q R_{pq}a_q^\dagger$ with the unitary matrix $R=e^{-{\kappa}}$. See the appendices of ref.~\citenum{Sokolov2020quantum, kottmann2022molecular} for details and applications, as well as  refs.~\citenum{mizukami2020,yalouz2021stateaveraged, kottmann2022optimized}.
\subsubsection{Clifford circuits}\label{subsec:clifford-circuits}
We now have a recipe for folding different kinds of circuits. Next, we explore the types of circuits that facilitate the recovery of correlation energy lost through the clustering of states.
Using quantum-chemical encoding (see e.g. reviews~\cite{cao2019quantum,mcardle2020quantum}), a molecular Hamiltonian can be written as a linear combination of Pauli strings 
\begin{equation}\label{eq:hamaslincombofpaulis}
H=\sum_\balpha c_\balpha P_\balpha = \sum_\balpha c_\balpha
\sigma_{\alpha_1} \otimes \cdots \otimes \sigma_{\alpha_{N_q}} \ ,
\end{equation}
where $\balpha = (\alpha_1, \ldots, \alpha_{N_q})$ and $\sigma_{\alpha_j} \in \{\identity, X, Y, Z \}$ is a Pauli matrix acting on site $\alpha_j$.
We transform the original Hamiltonian as described \review{in Section~\ref{subsubsec:folding}}{\hyperlink{subsubsec:folding}{above}} and measure expectation value of 
\begin{equation}
    \Tilde{H} = \clascirc^\dagger H \clascirc = \sum_\balpha c_\balpha \clascirc^\dagger P_\balpha \clascirc.
\end{equation}
The complexity of the measurement depends on the cardinality of the transformed Hamiltonian, i.e., the number of distinct Pauli strings in $\Tilde{H}$ that require independent measurement.

At this point, we notice that the cardinality of the transformed Hamiltonian can significantly grow, depending on the structure of $\clascirc$.
For single-qubit gates $\clascirc \in SU(2)$,  one can write 
\begin{equation}\label{eq:su2-identity}
  \clascirc = \exp(\imag \tau \Vec{n}\cdot\Vec{\sigma}) = \cos(\tau) \identity + \imag \sin(\tau) \Vec{n}\cdot\Vec{\sigma}, 
\end{equation}
where $\norm{\vec{n}} = 1,$ $\vec{\sigma}=(X,Y,Z)$. 
We further note that the Pauli operators (together with phase factors, see \cref{eq:pauli-group}) form a group. On one qubit, this means that due to the closure property, transformation of a Pauli operator by a single-qubit operation as in \cref{eq:su2-identity} can only produce terms with $\{\identity, X,Y,Z\}$, up to the phase factors. Thus for each single-qubit operator that transforms a Pauli-string in the Hamiltonian as in \cref{eq:hamaslincombofpaulis}, the number of terms grows at most by a factor of 4. 
Then, for a system of $\numqubits$ qubits, a virtual circuit consisting of only single-qubit gates leads to a cardinality of at most $\lvert \tilde{H} \rvert \le \order{4^\numqubits\lvert H \rvert}$~\cite{biamonte2021universal}. This bound holds as well for multi-qubit gates.

Thus, dressing the Hamiltonian with general gates is very costly and should be avoided. As an example, we show the increase in number of terms for folding a UCCSD circuit
into the Hamiltonians for H$_2$, BeH$_2$ and N$_2$ in \cref{fig:folding-uccsd-terms-increase}. 
The cost-effective solution is given by gates from the Clifford group $\cliffordgroup_n$, which is defined as the subgroup of the unitary group that normalizes the Pauli group. That is, the set of unitary operations that keep the Pauli group
\begin{align}\label{eq:pauli-group}
  \Pi_n = \Big\{ e^{\imag \theta \frac{\pi}{2}}\sigma_{0}\otimes\cdots\otimes\sigma_{n}:~& \theta\in\{0,1,2,3\},  \\ &\sigma_{j}\in\{\identity,X,Y,Z\}\Big\}  \nonumber
\end{align}
invariant under similarity transformations~\cite{tolar2018clifford}, i.e. 
\begin{equation}
  \cliffordgroup_n = \{ U :  U^\dagger \sigma U = \sigma'; \sigma,\sigma'\in \Pi_n\}.
\end{equation}
Hence for $\clascirc\in\cliffordgroup_n$, the cardinality of a Hamiltonian over $n$ qubits remains unchanged.  

However, for Clifford gates only, we just introduce a certain, restricted type of entanglement between the clusters as Clifford circuits are not universal~\cite{gottesman1998theory,jozsa2013classical}. This might already prove useful, but ultimately we considering \textit{near-Clifford} circuits, so that we can control the additional cost and take benefit from more general transformations. Such circuits are composed mostly out of Clifford gates, however, also contain a limited number of non-Clifford gates in order to boost expressiblity. 

ref.~\citenum{lang2020unitary} also has investigated the growth of Hamiltonian terms for certain non-Clifford circuits as the qubit coupled cluster (QCC) approach can be formulated as a folding of the Hamiltonian. They found that, while exponential increase cannot be avoided, the coefficient $c$ for an increase that goes as $c^N$, is smaller for so-called involutory linear combinations of entanglers that lead to a involution as opposed to standard QCC entanglers. 
More recently, ref.~\citenum{lang2022growth} extended this work by exploiting the structure of the Clifford group to build a Hamiltonian that is folded by so-called normalizer Pauli strings rather than qubit excitations. Within their approach, they circumvent the complications that come with a discrete optimization problem, however, \review{}{they} do not address a Clifford-only circuit, which is the main focus of this work.

\subsection{Optimizing Clifford circuits using a genetic algorithm}\hypertarget{subsec:optimizing-cliffords}{}

In this work, optimization refers to the process of minimizing the energy expectation value by optimizing the parameters in the circuits in the following sense: For the clusters $U_j(\theta)$ with fixed architecture, the parameters $\theta$ are varied. At the same time, both structure and parameters need to be adjusted for the near-Clifford circuits, which are supposed to account lost correlation due to the enforced product ansatz in the $U_j$'s. 
For the case when $M\in\cliffordgroup_{\numqubits}$, there are no continuous parameters available that would allow an adaptive scheme such as in the family of ADAPT-VQE methods~\cite{grimsley2019adaptive,tang2021qubit} and qubit coupled cluster techniques~\cite{ryabinkin2018qubit,ryabinkin2020iterative} as the Clifford group only forms a discrete set. One could think of Bayesian mappings from the discrete gate set to a continuous parameter such as  in ref.~\citenum{burton2022exact} or a simple simulated annealing strategy as in ref.~\citenum{ravi2022cafqa}. In our tests, a simulated annealing strategy \review{}{was} surpassed by a genetic algorithm, which is what we use and will describe in the following.

The genetic algorithm we choose resorts to an operator pool, which it uses as a reference set for the discrete optimization problem. Elements of the operator pool are used to modify the circuit by adding new operators and modifying existing gates in $\clascirc$. 

The operator pool we use can be found in \cref{tab:operator-pool-clifford}. 
Aside from the generators of the Clifford group ($H, S, C_X$), we also include the Pauli gates, controlled Pauli gates, a SWAP gate to mimic permutations. These have proven helpful in ref.~\citenum{tkachenko2021correlation} to reorganize the correlation pattern in the state, so it is easier to describe with a cluster product ansatz. We also include gate blueprints that are inspired by fermionic excitations~\cite{anand2021quantum}, where we fix the angle of the freely parametrized $R_Z$-rotation in the center aiming to obtain an approximate representation in form of a Clifford gate so that it is either one of $S, Z, S^\dagger$. These are supposed to mimic gate sequences that have proven useful in finding ground-states for quantum chemistry. Extending the gate pool may aid to speed up optimization, as these sequences can be picked from the pool and do not need to be found via optimization. 
Next, we explain the optimization procedure to find a Clifford circuit $M$, which is also described in algorithm~\ref{algorithm:procedure}.

\begin{figure*}
    \centering
    \includegraphics[width=\linewidth]{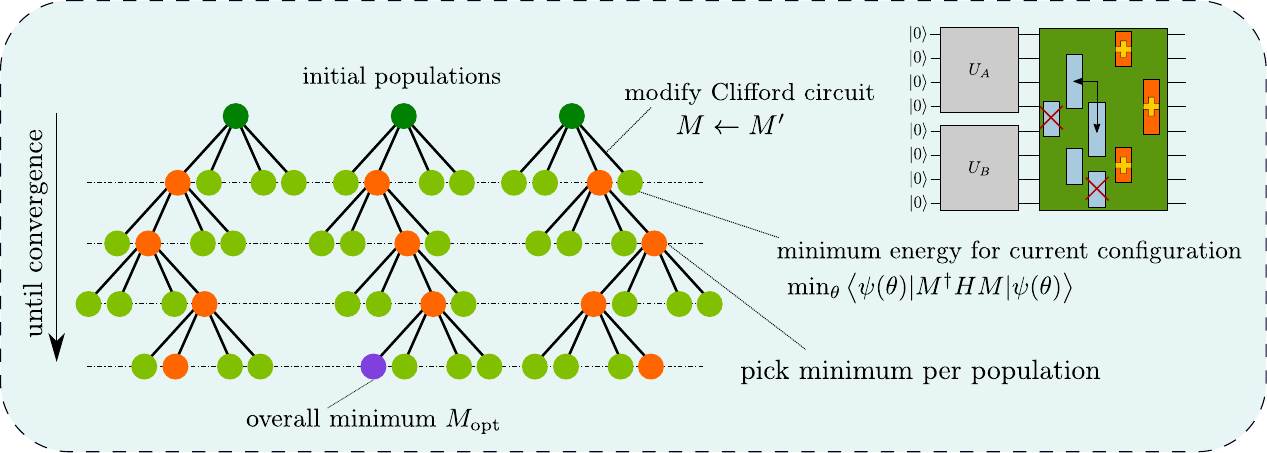}
    \caption{Outline of the genetic algorithm used to assemble the classical circuit $M$. Based on some initial circuits (initial populations) made up by random gates drawn from the operator pool in Tab.~\ref{tab:operator-pool-clifford}, random changes of this circuit in form of addition, deletion and change of gates as shown in the top-right corner, are proposed. The best change per circuit is used as a reference for the next iteration as long as this change does not perform worse than the reference method; this procedure is repeated until convergence or until a maximum number of iteration is reached.}
    \label{fig:genetic-outline}
\end{figure*}

The general procedure to generate the classical circuit is outlined in \cref{fig:genetic-outline}. First, we generate a set of initial populations by randomly sampling circuits built from the operator pool. Any initial gates that do not have support over multiple clusters can be seen as an extension/modification of the cluster circuits; thus, one can choose to avoid them in the optimization procedure. We discuss this further \review{in Section~\ref{subsec:circuits}}{\hyperlink{subsec:circuits}{}}.

Based on the initial circuits per population, we impose a set of modifications, drawn randomly from the set of allowed operations. These are comprised of the addition of gates from the pool, exchanging specific gates with others or re-arranging them and deleting gates. We choose a scheduling so that each modification occurs with a certain probability; e.g., we found that favoring changing over deletion and addition is advantageous, while during early iterations the rate of additions should initially be higher. 

\begin{algorithm}
\caption{Architecture search for $M$}
\label{algorithm:procedure}
    \begin{algorithmic}[1]
        \State \textbf{Input} operator pool, reference method / circuits, 
         number of populations, offsprings per population and maximum iterations
        \ForAll{ populations }
        \State compute reference energy
        \State sample initial circuits $M$
        \State optimize reference to compute energy with initial circuits
        \EndFor
        \While{ not converged }
            \ForAll{ populations }
                \ForAll{ offsprings per population }
                    \State propose modification to $M$
                    \State optimize reference to compute energy  
                \EndFor
                \State pick offspring to continue with based with probability $e^{-\frac{\Delta E}{T}}$, update $T$
            \EndFor
        \EndWhile
        \State \textbf{Output} circuit $M$, optimal reference, approximation to ground-state energy
    \end{algorithmic}
\end{algorithm}

Then, for each population, we allow a number of off-springs/children with independent modifications. We pick the child with the lowest energy to continue the search. After a set number of iterations or when a convergence criterion is fulfilled -- here, the change in energy between iterations -- we end the optimization and pick the minimum over all populations as an approximation to the optimal solution. Within the search, we make use of an additional ``patience'' hyperparameter -- this parameter is increased whenever the modifications does not lead to an improvement in energy. When the patience parameter reaches a certain value, the associated population ``frustrates'' and re-sets its circuit instructions to a previous stage. We further allow updates such as in simulated annealing, where we pick the best offspring per population with probability $e^{-\frac{\Delta E}{T}}$ and decrease temperature $T$ over the course of iterations. As a result, the best improvement in energy is picked most likely but there is still a chance that even an increase in energy would be allowed to avoid local minima; however, we do not allow increases in energy beyond the minmium energy that was found for the employed reference method $U$ with $M=\identity$.

After each such optimization step, the parameters in the cluster circuit $U_j(\theta)$ are adjusted.
For the case of Clifford circuits, one may consider techniques proposed in refs.~\citenum{fagan2019optimising,richter2022transport} to further optimize the depth of the resulting circuits.

\begin{table}[t]
    \centering
    \begin{adjustbox}{width=.95\linewidth}
    \begin{tabular}{cc}
         Pool of Clifford gates & $ \{ H, S, X, Y, Z, C_X, C_Y, C
_Z, \textrm{SWAP}, \textrm{Exc}_1, \textrm{Exc}_2  \}     $    \\
         & \\
         $\textrm{Exc}_1$ &
         \begin{adjustbox}{width=.2\textwidth}
         \begin{tikzcd}
            & \gate{} & \ctrl{1} &  \qw & \ctrl{1}  & \gate{}  &\qw   \\
            & \gate{} & \targ{} &  \gate{\textrm{\Large Rot}} & \targ{} & \gate{} &  \qw 
        \end{tikzcd}  
        \end{adjustbox}
         \\
         & \\
         $\textrm{Exc}_2$ & 
         \begin{adjustbox}{width=.25\textwidth}
         \begin{tikzcd}
            &\gate{} &\ctrl{1}&\qw     & \qw     & \qw                 &\qw     &\qw     &\ctrl{1} &\gate{}&\qw \\
            &\gate{} &\targ{} &\ctrl{1}& \qw     & \qw                 &\qw     &\ctrl{1}&\targ{}  &\gate{}&\qw \\ 
            &\gate{} &\qw     &\targ{} & \ctrl{1}& \qw                 &\ctrl{1}&\targ{} &\qw      &\gate{}&\qw \\
            &\gate{} &\qw     &\qw     & \targ{} & \gate{\textrm{\Large Rot}} &\targ{} &\qw     &\qw      &\gate{}&\qw
        \end{tikzcd}  
         \end{adjustbox}
         \\
         & \\
         $\textrm{Rot}\in\{S, Z, S^\dagger\}$, &\begin{tikzcd} & \gate{}&\qw \end{tikzcd} $\in \{R_X(\pi/2), R_Y(\pi/2)\}$
    \end{tabular}
    \end{adjustbox}
    \caption{Operator pool used for Clifford optimization including excitation-like gates $\textrm{Exc}_1,\textrm{Exc}_2$. These consist of basis transformations, CNOT ladders and a central $Z$-rotation, which implements operations $e^{-\imag \frac{\theta}{2} \sigma_1 \sigma_2 \cdots \sigma_{\numqubits}}$, where the basis transformations allow $\sigma_i$ to be any of $\{X,Y,Z\}$. Further, Rot$~\in\{S,Z,S^\dagger\}$, $\theta\in\{\frac{\pi}{2}, \pi, \frac{3\pi}{2}\}$. \review{}{Comment: The orange in the circuits will disappear in the version without highlighting the changes.}}
    \label{tab:operator-pool-clifford}
\end{table}

\subsection{Near-Clifford circuits to increase expressibility}
It is well-known that circuits made up from Clifford gates only are not universal \cite{gottesman1998theory}. We aim to counter-act this by adding few non-Clifford gates to the circuit $\clascirc$. 
However, we also know that including more and more non-Cliffords likely leads to an immense increase in number of terms in the Hamiltonian, making measurement intractable\review{ (cf. Section~\ref{subsec:clifford-circuits})}{}. Thus what we understand as a near-Clifford circuit is a circuit that results in a ``reasonable'' increase in the sense that $\lvert \Tilde{H} \rvert \le C \lvert H \rvert$ for some moderately-sized constant $C$. In particular, we want to avoid a regime with exponential increase. An easy choice whose capabilities we will explore later on is modifying a single Clifford gate towards its parametrized version, i.e. for a Clifford gate $U_\text{Cliff}$ with generator form $U_{\text{Cliff}} = \exp(-\imag \theta_\text{Cliff} G)$ and fixed $\theta_\text{Cliff}$, we use $U(\theta) = \exp(-\imag \theta G)$ with a free parameter that can be used for optimization.
Hence, for our experiments here, we will loop over all Clifford gates within the virtual circuits $M$, except for controlled gates and SWAP, and replace them with parametrized versions. We outline this strategy in algorithm~\ref{algorithm:procedure-tau}.
For example, we substitute $X \to R_X(\tau)$
or $\textrm{Exc}_{1;2}$ (cp. with Tab.~\ref{tab:operator-pool-clifford}) receive a $R_Z(\tau)$ as center gate. 


\begin{algorithm}
\caption{Optimization of near-Clifford $\clascirc$ with one single-qubit non-Clifford gate}
\label{algorithm:procedure-tau}
    \begin{algorithmic}[1]
        \State \textbf{Input} non-Clifford version of operator pool, reference method / circuits, 
         optimized Clifford circuit $M$ and maximum iterations
        \ForAll{ (single-qubit) Clifford gates $V$ in $M$ }
        \State replace $V$ with its non-Clifford version $g(\tau)$ and initial parameter set to correspond to Clifford
            \While{ not converged }
                \State optimize reference
                \State optimize parameters in $V(\tau)$ with only few iterations 
            \EndWhile
            \State re-update reference and then $\tau$ with smaller tolerance/more iterations
        \EndFor
        \State set $V(\tau)$ with best performance as non-Clifford
        \State \textbf{Output} circuit $\clascirc$, optimal parameter $\tau^\star$, optimal reference and approximation to ground-state energy
    \end{algorithmic}
\end{algorithm}

\section{Results}\hypertarget{sec:results}{}
We will demonstrate our proposed technique on a set of small molecular examples to validate capabilities of Clifford and near-Clifford circuits to re-gain correlation due to the assumed product state. For the hydrogen molecule, we choose a basis set of Gaussians, whereas for beryllium hydride and the nitrogen molecule, we use a basis of pair-natural orbitals generated by multi-resolution analysis~\cite{kottmann2021reducing,kottmann2022optimized}, as described before.

\review{}{The simulations are based on a pre-optimized product state
\begin{equation}\label{eq:reference-state}
    \ket{\textrm{ref}} = \big(\prod_{j=1}^{\numclusters} U_j\big) \ket{0}
\end{equation}
that we call reference method. That is, ``reference'' within this study decribes a product state over chosen qubit clusters and not describe the Hartree-Fock state.}
We choose the following reference method for all our systems: We use a power method-type procedure that enforces a product state in the wavefunction to find the ground-state. A brief overview of this method can be found \hyperlink{appendix:reference-method}{in the SI}. 

\subsection{Simulation setup}\label{subsec:simsetup}
Simulation code was written with the \textsc{tequila}~\cite{kottmann2021tequila} framework using the \textsc{qulacs} simulator~\cite{suzuki2021qulacs} as simulation backend for the quantum circuits and the automatically differentiable framework of~\cite{kottmann2021feasible}. Direct basis determination via pair-natural orbitals was performed with \textsc{madness}~\cite{harrison2016madness} as described in ref.~\citenum{kottmann2021reducing} with the diagonal approximation described in ref.~\citenum{kottmann2022optimized} and the MP2-PNO implementation of ref.~\citenum{kottmann2020direct} without cusp regularization. The latter interfaces mean-field implementations described in refs.~\citenum{harrison2004multiresolution, bischoff2014regularizing}. The Jordan-Wigner encoding corresponds to the implementation in \textsc{OpenFermion}~\cite{mcclean2020openfermion}. Orbitals in standard basis sets were computed with \textsc{psi4}~\cite{smith2020psi4}\review{}{while orbital optimization was carried out in \textsc{pyscf}~\cite{sun2020recent,zhang2022differentiable}}.

For the genetic algorithm as outlined in \cref{fig:genetic-outline}, we choose the number of populations between 10--15 and the number of offsprings per population as 8--10. The convergence criterion for the reference method is set to $10^{-5}$, and the maximum number of iterations (tree depth in \cref{fig:genetic-outline}) is set between 10--25. \review{}{We observed that the performance of the algorithm is quite insensitive to these parameters (around $10^{-4}$ milli-Hartrees), which motivates our choice, cp. to \cref{fig:parameter-search-genetic} \hyperlink{appendix:parameter-choice}{the SI}}.  

For all the following results, we show the error of various employed methods with respect to the exact ground state (Full Configuration Interaction) within the same basis set. The reference state as in~\cref{eq:reference-state}\review{, which produces and optimizes a product state via $\prod_j U_j$, we use}{ is obtained following the procedure in \hyperlink{appendix:reference-method}{in the SI}}. 
With this as a starting point, we assemble a virtual circuit $M$ built from Clifford gates only (``Reference + Cliffords'') and subsequently search over all Clifford gates and check if the energy can be improved by optimizing their parameter when making them parametrized (``Reference + Near-Clifford circuit'').

\subsection*{Hydrogen molecule}
\begin{figure}%
\centering%
    \centering%
    \includegraphics[width=.7\linewidth]{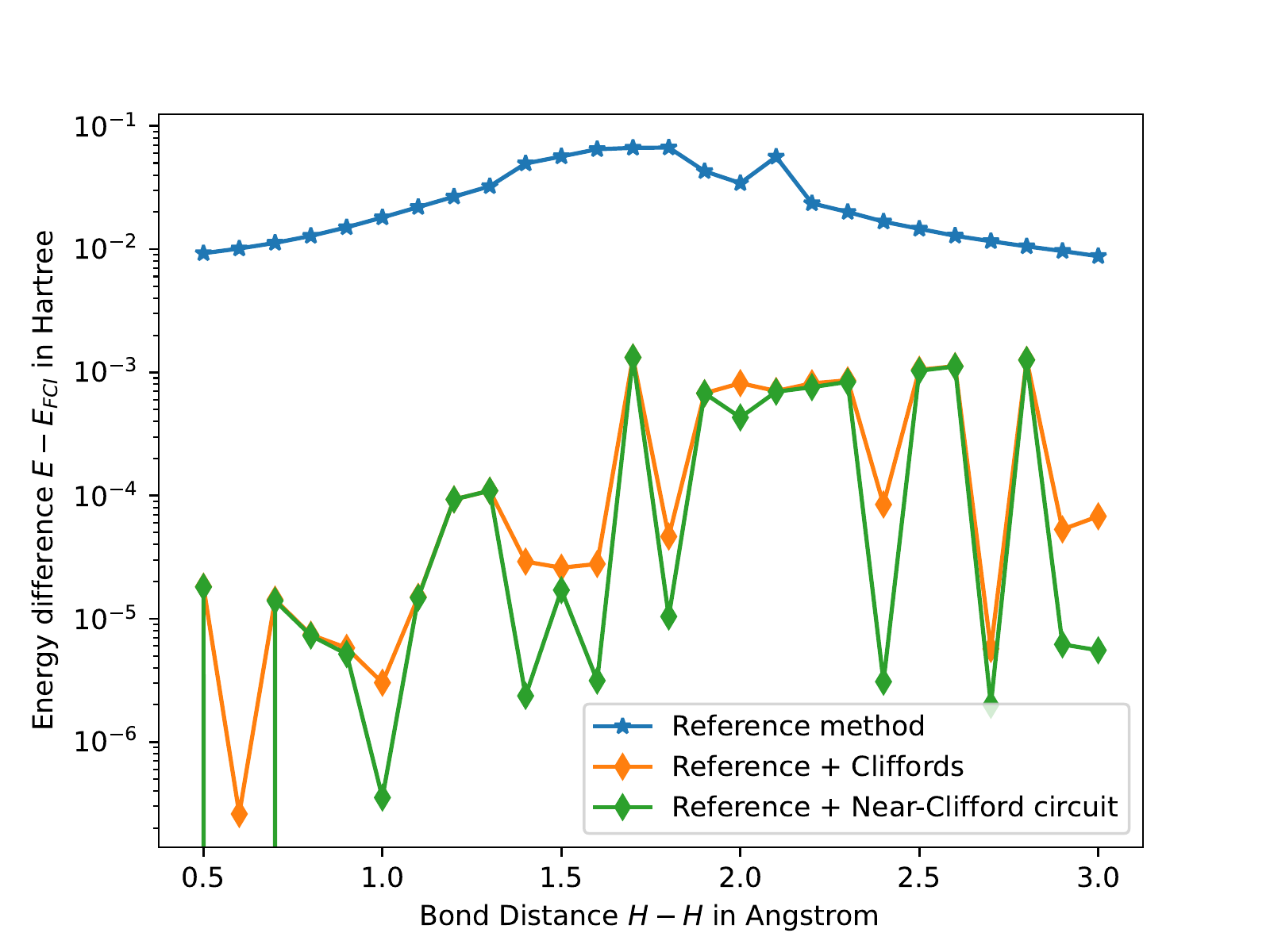}%
    \caption{Simulating the ground-state of H$_2$ in the 6-31G basis set using an arbitrary active space of six qubits; here, we also use an uninformed clustering, where the first set of three and the second set of three qubits are enforced into a separable state.%
    \review{}{``Reference method'' refers to the power-method based reference from \cref{eq:reference-state} or \cref{appendix:reference-method} in the SI}.}
    \label{fig:h2-pes}%
\end{figure}

We use the 6-31G basis set for H$_2$, where we choose an active space of six spin-orbitals to obtain a six-qubit toy example. As a reference, we enforce a product state \review{over the ``first'' and the ``last'' three orbitals}{of overall six qubits, made up by the product over a state containing the ``first'' and another one containing the ``last'' three orbitals}. Here, first and last corresponds to the ordering with respect to orbital energies. As showcased in \cref{fig:h2-pes}, this approach has a rather high error, of the order of hundreds of milli-Hartree. However, addition of a virtual Clifford circuit enables to suppress the error to the order of milli-Hartrees.
In the case of small bond distances, the reduction is as much as three orders of magnitude. For large bond distances, the improvement by adding additional entanglement is lower but still accounts for roughly one order of magnitude.
This is because the true state is close to a state preparable by a pure Clifford circuit supported only on the first four qubits as explained in ref.~\citenum{kottmann2022molecular}.
To sum up, we observe significant improvement for Clifford-gates only. While the addition of freely parametrized non-Clifford gates does not reliably lower the energy, if successful, it can yield up to another order of magnitude in accuracy.  

\subsection*{Nitrogen molecule}
\begin{figure}%
    \centering%
    \includegraphics[width=.7\linewidth]{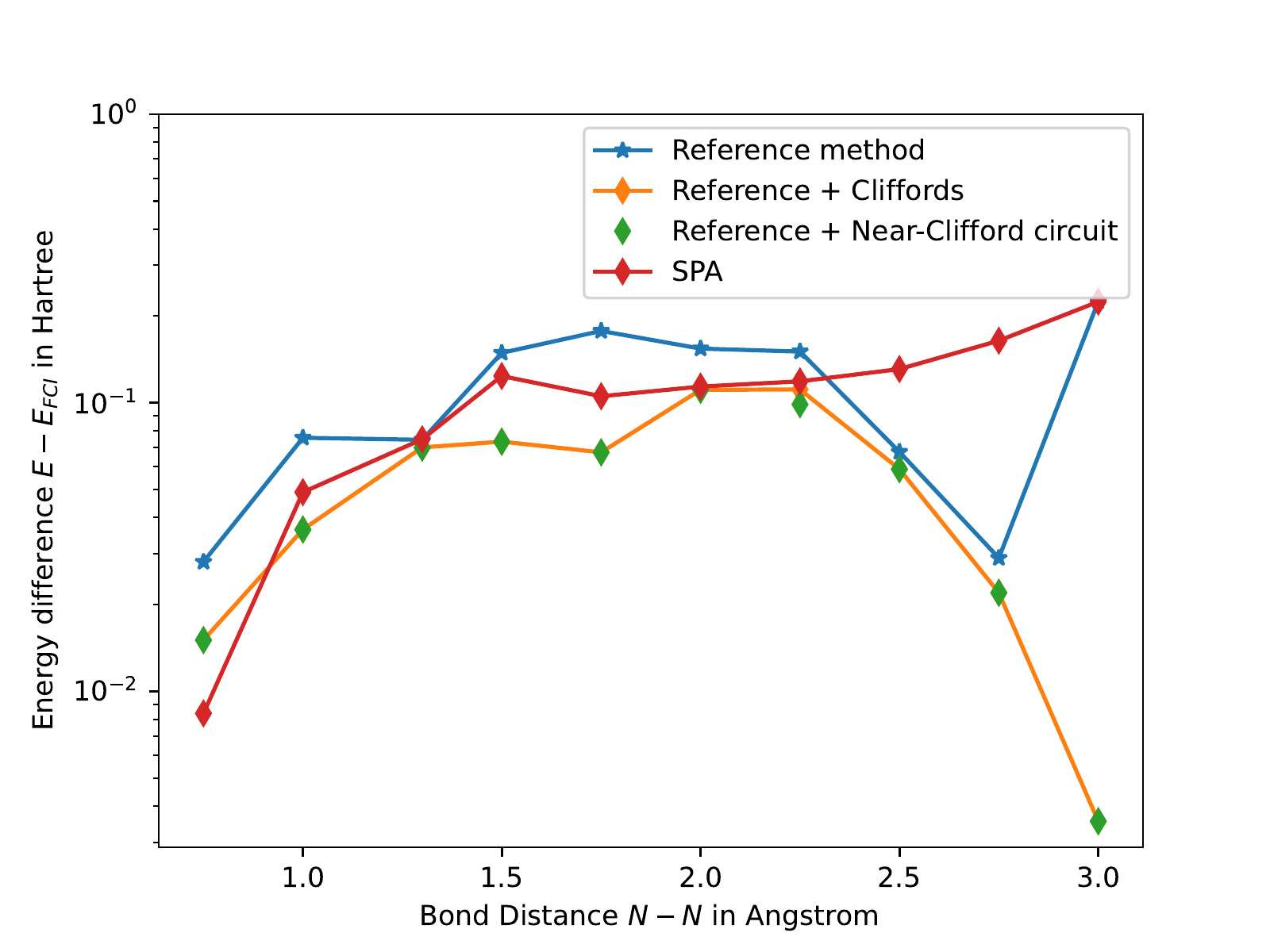}%
    \caption{N$_2$ with two SPA pairs as clusters in frozen-core approximation, 12 qubit simulation. \review{}{``Reference method'' refers to the power-method based reference from \cref{eq:reference-state} or \cref{appendix:reference-method} in the SI}.}%
    \label{fig:n2-pes}%
\end{figure}

For the N$_2$ molecule, as depicted in \cref{fig:n2-pes}, the SPA outperforms our product state reference method in most regimes\review{}{, while still not coming close to FCI level}.
\review{However, for long bond distances, we can observe significant improvement.}{} The SPA performs \review{}{worse} here due to the surrogate model that is used to generate the PNOs which are used as a basis. They are generated using second-order Møller-Plesset perturbation theory \review{as a surrogate model}{} and thus the PNOs inherit \review{}{its} shortcomings\review{of the surrogate model}{}~\cite{kottmann2020direct,kottmann2021reducing,schleich2022improving}. 
\review{}{Even more powerful models, such as $k$-UpCCGSD, struggle here~\cite{lee2018generalized,kottmann2022optimized}.
While our method makes use of the same PNO basis, it is, in contrast to the SPA, able to recover a part of this error.}{} 

Upon addition of a virtual Clifford circuit, we again observe a consistent improvement. 
\review{It is noticeable that the SPA is still better for the lowest bond-length considered and that our method seems to perform better for short and large bond lengths.}{
It is noticeable that for short bond lengths, SPA and our method overall perform quite comparably. For longer bond lengths, despite using the same PNO basis as SPA, it seems to be able to recover part of the error due to the basis.} 
Furthermore, we found that parametrizing a single Clifford gate of the virtual circuit for the most part improves the energies compared to the Clifford circuit only barely. \review{For this system, the genetic algorithm also showed rather unfavorable convergence behavior and tends to take very long time to achieve convergence.}{As the parameter space for N$_2$ is already considerably large, convergence of the genetic algorithm for the Clifford-circuit already takes rather long. The increase in complexity by adding even a single non-Clifford gate already makes it hard for the algorithm to find optimal configurations.}  

\subsection*{Beryllium hydride}
We further consider BeH$_2$ as a system of interest \review{for comparability}{in comparison} with results from ref.~\citenum{kottmann2022optimized}. We use a basis of eight qubits\review{}{ in a frozen-core approximation} using MRA-PNOs (PNOs determined by multi-resolution analysis), made up by two clusters \review{formed by}{of} two spatial orbitals based on each \review{}{bonding} electron pair. As expected, we can see in \cref{fig:beh2-spa} that the SPA-ansatz slightly outperforms our reference method.  
However, the addition of a virtual circuit made up by Cliffords only allows to improve beyond the SPA\review{ and this}{. This} can be further boosted by parametrizing one of the Clifford gates. 
We \review{further}{} note that the improvement seems to produce a somewhat constant offset\review{ --}{, while} the general behaviour of the error curve remains unchanged. 
This behaviour is similar to \review{standard}{} orbital optimization on top of the SPA reference (OO-SPA\review{, as done in ref.~)}{), as done in ref.~}\citenum{kottmann2022optimized}. 
However, we note that the improvement by the \review{found}{optimized} Clifford \review{}{circuits} and especially near-Clifford circuits is generally slightly higher. \review{This is despite that orbital optimizations are non-Clifford operations, but leave the complexity of the Hamiltonian unchanged.}{This is despite of the orbital optimizations being non-Clifford operations, which are typically expected to be more expressible.}  
We observed that the behaviour of the procedure is more favorable if power method-based reference method is used instead of the SPA as a reference. The outcome of the initial optimization is somewhat randomized and therefore has eased to find gate sequences that lower the energy in subsequent iterations. \review{Potentially, the SPA is stuck in a more stable local minimum than the iterative power-method based solver, which would make it harder to break free.}{We expect SPA to be stuck in a more stable local minimum as compared to the iterative power-method solver, as for SPA, our procedure consistently either aborted early or only led to  circuits outputting the same energy.} 
\review{}{Beyond that, we provide a discussion on  fidelity computations for optimization outcomes with eigenstates obtained by direct diagonalization of the Hamiltonian in the eight qubit basis in \hyperlink{appendix:fidelity-computation}{the SI}, \cref{fig:overlap-figs}.

}

We next investigate the consequences of \review{not utilizing the optimal clustering into electron pairs prior to the analysis}{using non-optimal clustering, i.e. clusters from different bonding electron pairs}. We mix pairs so that \review{one}{a} spin-orbital of one pair is exchanged with one the other pair. Results for this case can be found in \cref{fig:beh2-perm,fig:beh2-deltas} as well as that the reference method behaves significantly worse. However, this does not seem to influence the outcome of the computations that include a virtual circuit. This is likely due to the inclusion of SWAP gates into the operator pool, which are able to retrieve the optimal clustering by permuting orbitals~\cite{tkachenko2021correlation}. 
In fact, we notice here that the circuits built from Clifford gates only perform as well as the near-Clifford circuits for most geometries and performs slightly better as in the un-permuted case. Since the employed algorithm is random and the outcome of the genetic algorithm with a finite number of populations and offsprings has some variance, this is not surprising. 

\begin{figure}[tb]%
\begin{subfigure}{.7\linewidth}%
    \centering%
    \includegraphics[width=\linewidth]{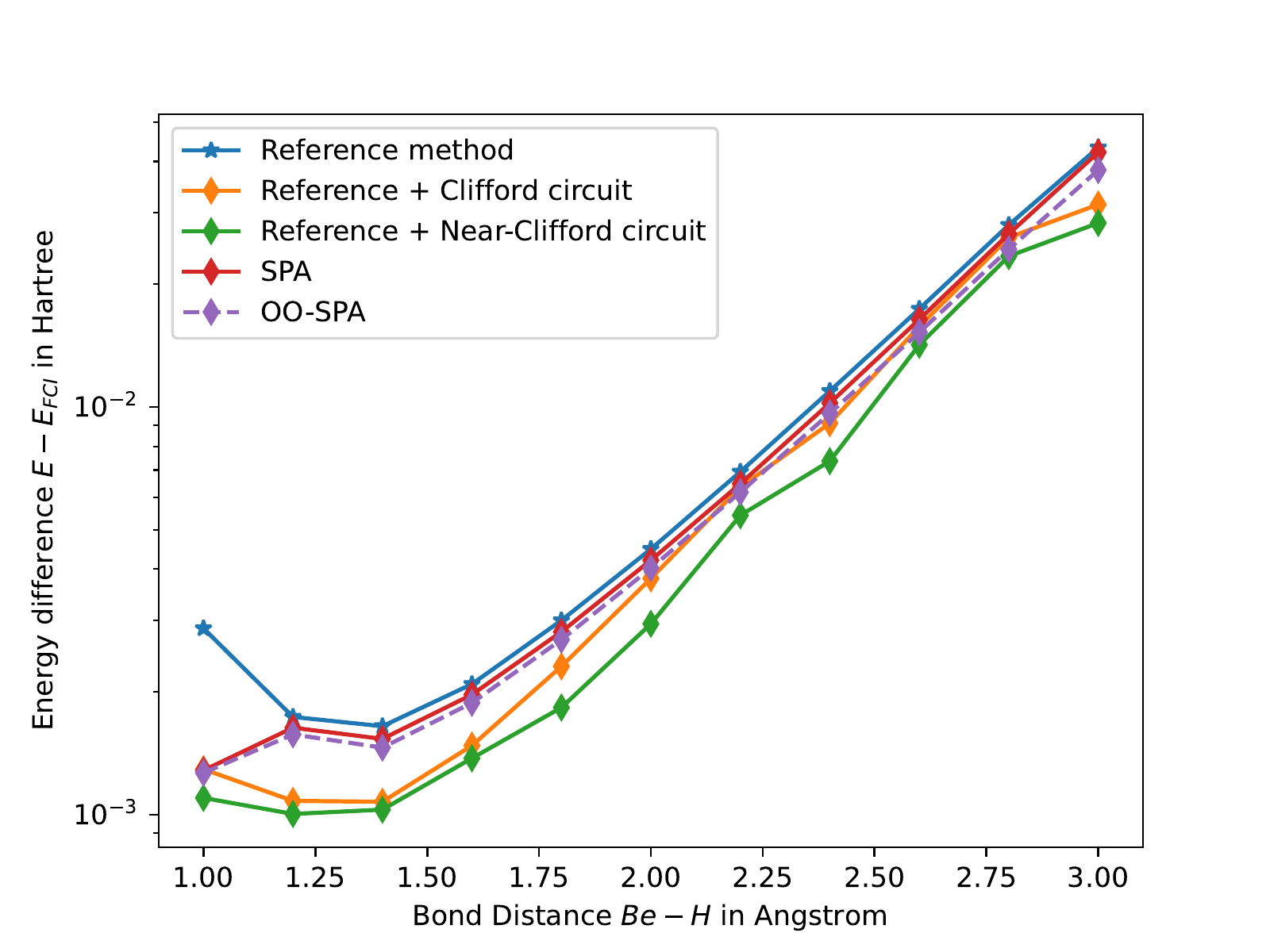}%
    \caption{BeH$_2$ with clusters according to electron pairs, 8 qubit simulation}%
    \label{fig:beh2-spa}%
\end{subfigure}%
\end{figure}%
\begin{figure}\ContinuedFloat%
\begin{subfigure}{.7\linewidth}%
    \centering%
    \includegraphics[width=\linewidth]{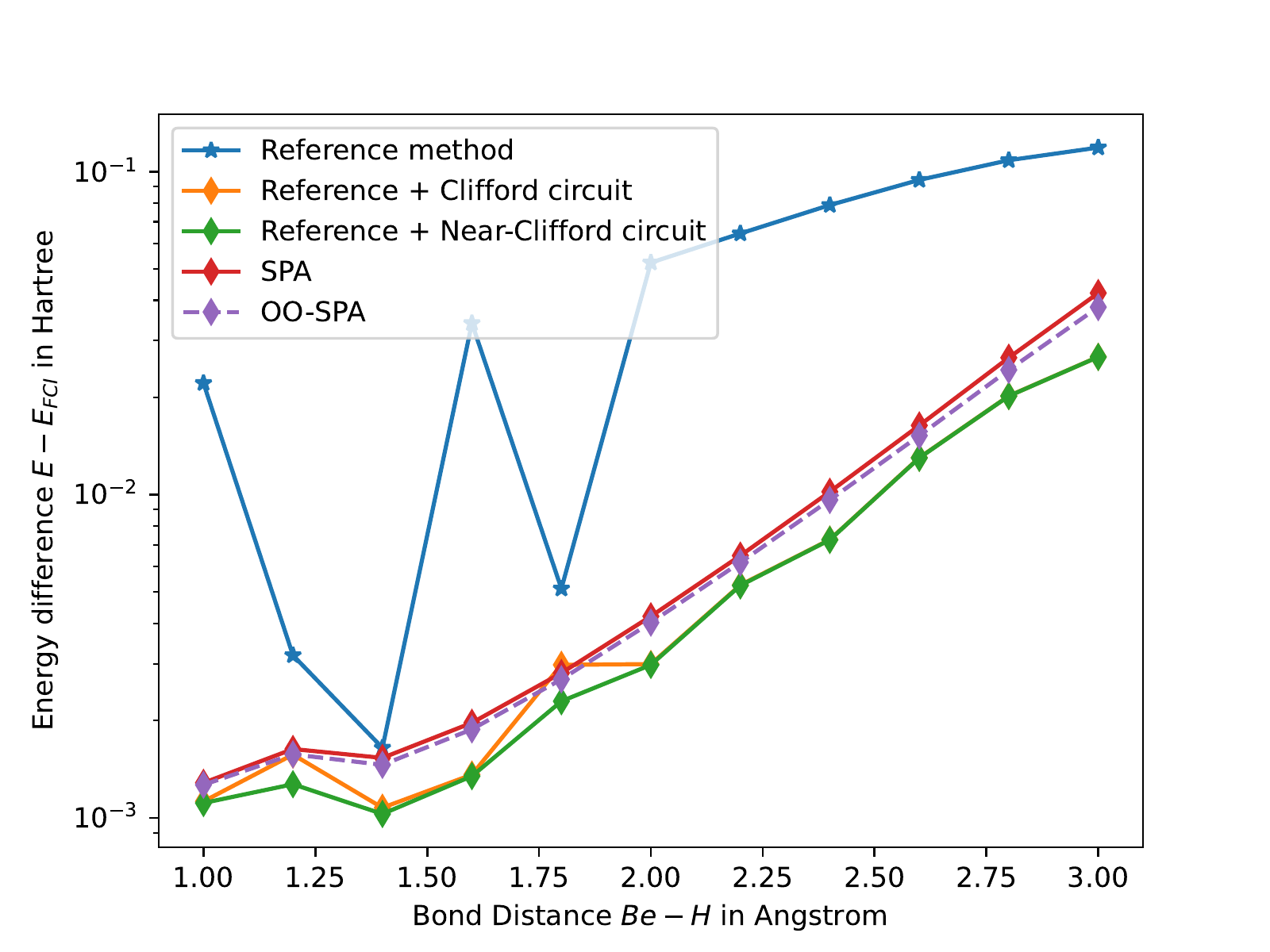}%
    \caption{BeH$_2$ with permuted reference clusters, also see \cref{fig:beh2-deltas}.}%
    \label{fig:beh2-perm}%
\end{subfigure}
\end{figure}
\begin{figure}\ContinuedFloat
\begin{subfigure}{.7\linewidth}%
    \centering%
    \includegraphics[width=\linewidth]{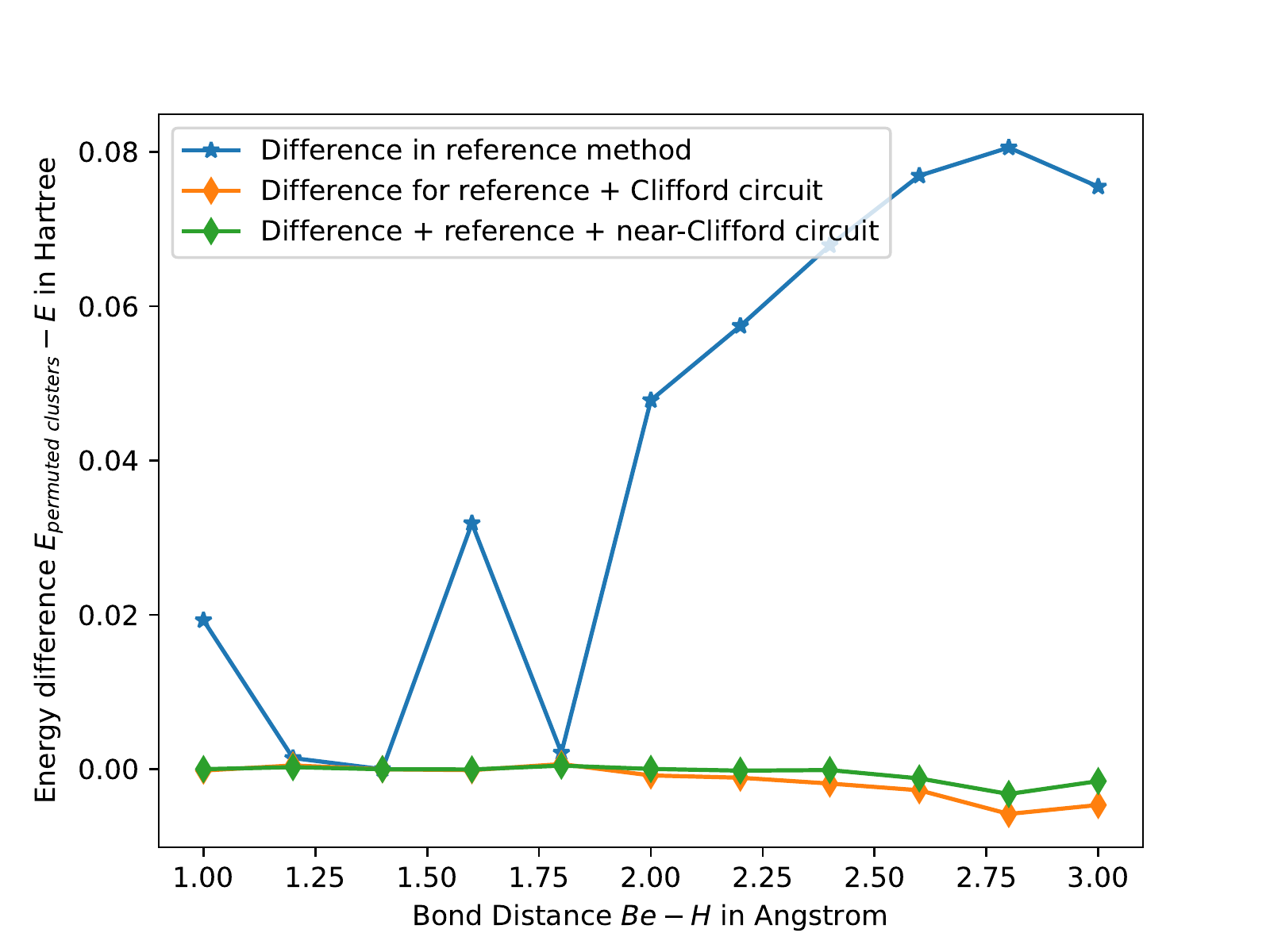}%
    \caption{Comparison of BeH$_2$ with pair-wise and modified clusters, showing energy difference between permuted clusters and \review{clusers}{clusters} by electron pairs. For a configuration between clusters $A$ and $B$ as $(000_A, 111_B)$, where 0,1 shows association to an electron pair, we permute to $(100_A, 011_B)$. }%
    \label{fig:beh2-deltas}%
\end{subfigure}%
\caption{Errors of optimization results with respect to FCI in corresponding bases. \review{}{``Reference method'' refers to the power-method based reference from \cref{eq:reference-state} or \cref{appendix:reference-method} in the SI}.}%
\end{figure}%

\subsection{Increase in the number of Hamiltonian terms}

{
\begin{figure}%
     \centering%
\begin{subfigure}{\linewidth}
     \centering
     \includegraphics[width=.6\linewidth]{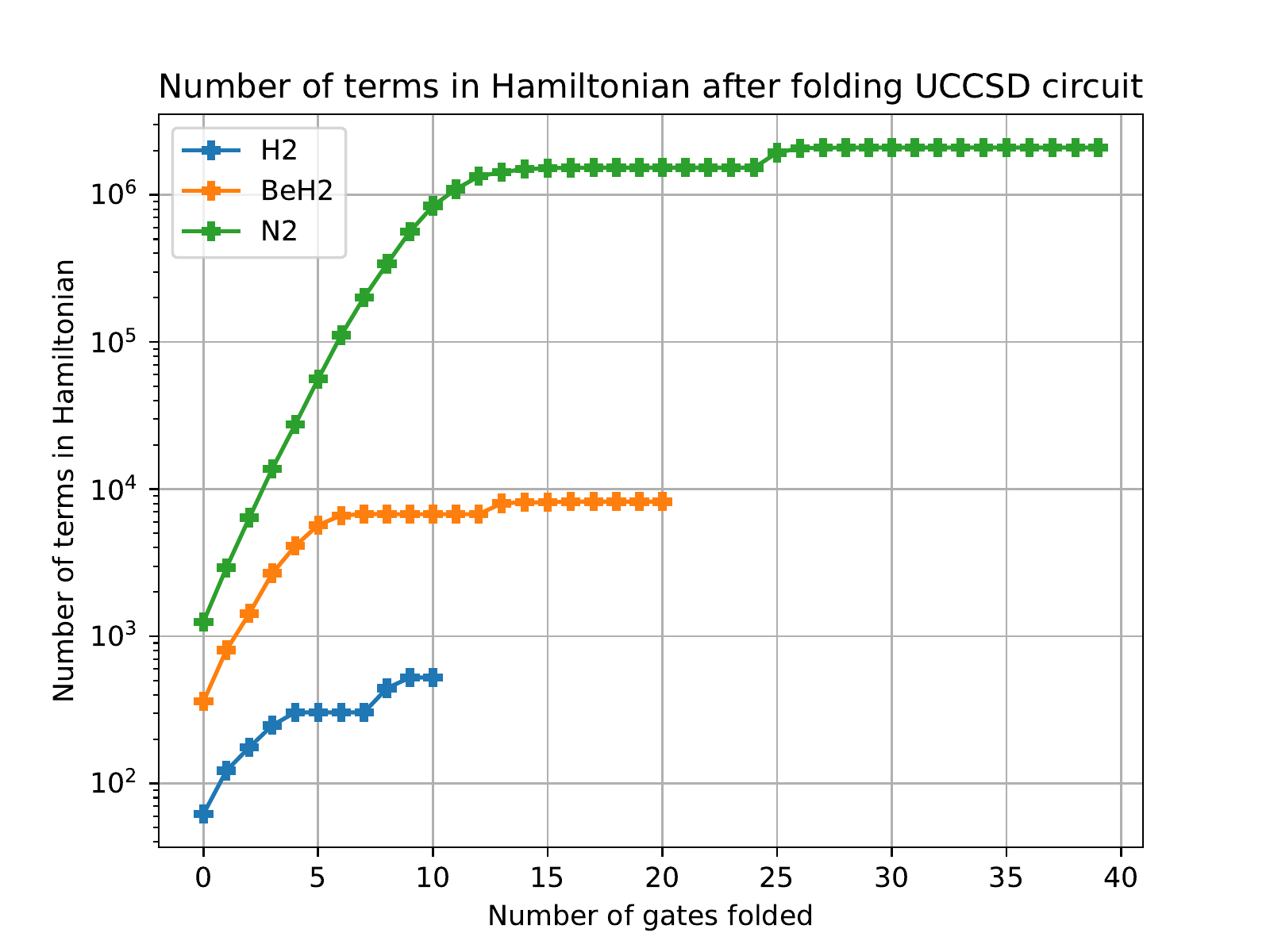}
     \caption{Increase in number of terms in the Hamiltonian for H$_2$, BeH$_2$ and N$_2$ molecules when folding a vanilla UCCSD circuit~\cite{anand2021quantum}. The horizontal lines without term increase correspond to Clifford gates.}
     \label{fig:folding-uccsd-terms-increase}
\end{subfigure}
\begin{subfigure}{\linewidth}
     \centering
    \includegraphics[width=\linewidth]{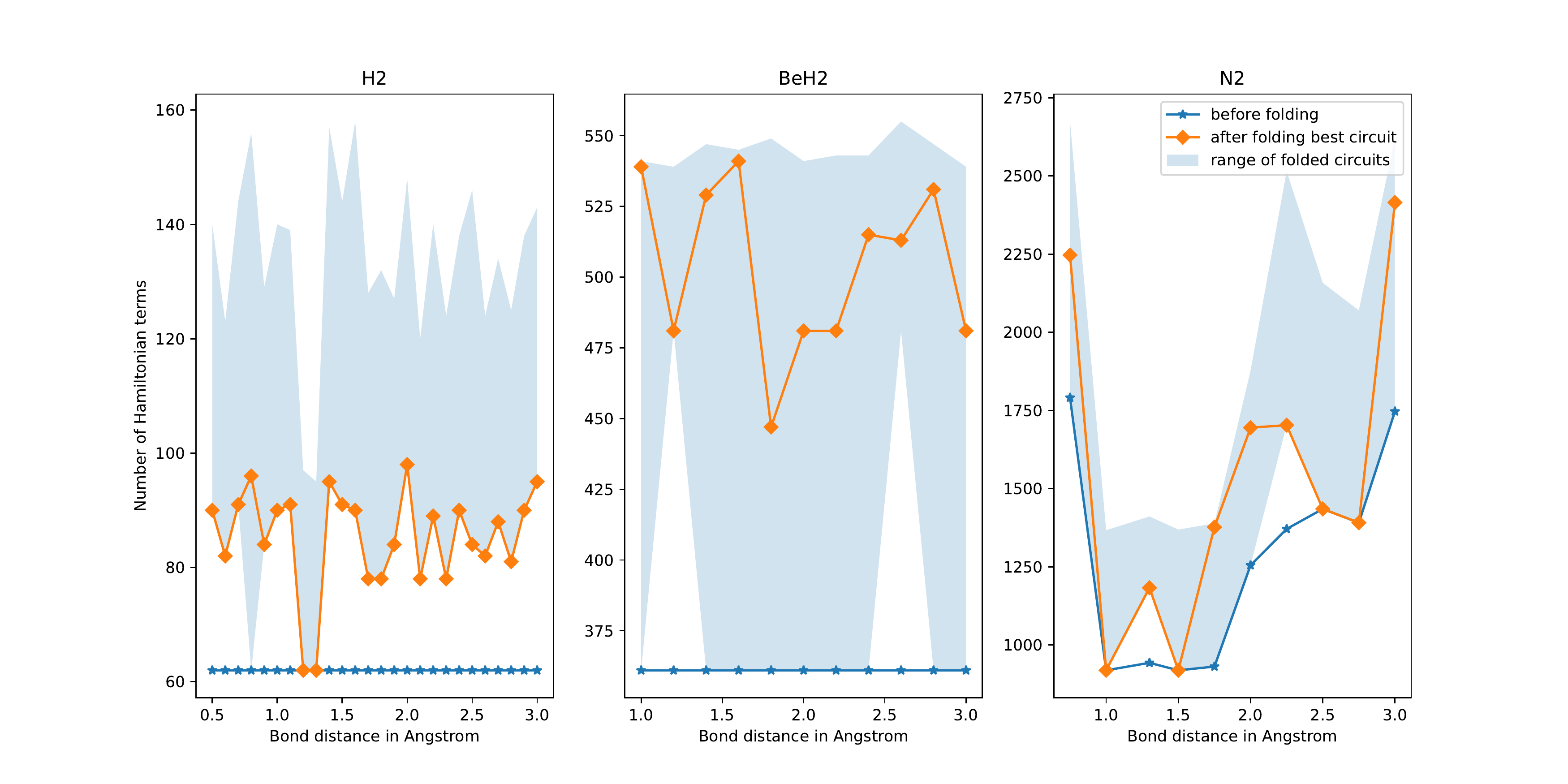}
    \caption{Increase of the number of terms in the Hamiltonian after folding the optimized circuits when using only one gate of the circuit as a parametrized non-Clifford gate. The range of folded circuits depicts the envelope of maximum- and minimum number of terms that are possible when parametrizing a single gate of the previously optimized Clifford circuit, assuming the parametrization makes them non-Clifford. Otherwise the minimum coincides with the number of terms before folding (blue line).
    The theoretically maximum number of terms are as follows. H$_2$: $\numqubits=6, 4^\numqubits=4096$; BeH$_2$: $\numqubits=8, 4^\numqubits = 65536$; N$_2$: $\numqubits=12, 4^{\numqubits} = 16777216$.}
    \label{fig:terms-increase-one-noncliff}
\end{subfigure}
\caption{Increase in Hamiltonian terms for non-Clifford (a) and near-Clifford (b) circuits}
\end{figure}
}

For the results discussed above, we show the increase of the number of terms in the Hamiltonian when folding the classical circuit in \cref{fig:terms-increase-one-noncliff}, compared to folding a full UCCSD circuit several non-Clifford gates in \cref{fig:folding-uccsd-terms-increase}. We show both the cardinality before ($\lvert H\rvert$) and after folding ($\lvert\Tilde{H}\rvert$). Additionally we demonstrate the range of cardinalities that can be accessed by exchanging one Clifford gate for another one in the previously optimized Clifford circuits for each geometry. As expected, the upper bound of $4^{\numqubits}$ times the previous complexity is respected. 

While it is clear that the measurement cost increases, we note that the increase when folding one gate should be manageable given that a single non-Clifford gate only increases the complexity by a constant factor. In fact, in the case of hydrogen, the cardinality after folding for the optimized circuit is almost at the lower bound of the range of circuits. However, this cannot be established as a trend for the larger systems. 

\subsection{Circuit structures}\hypertarget{subsec:circuits}{}
\begin{figure*}
    \centering
    \includegraphics[width=\linewidth]{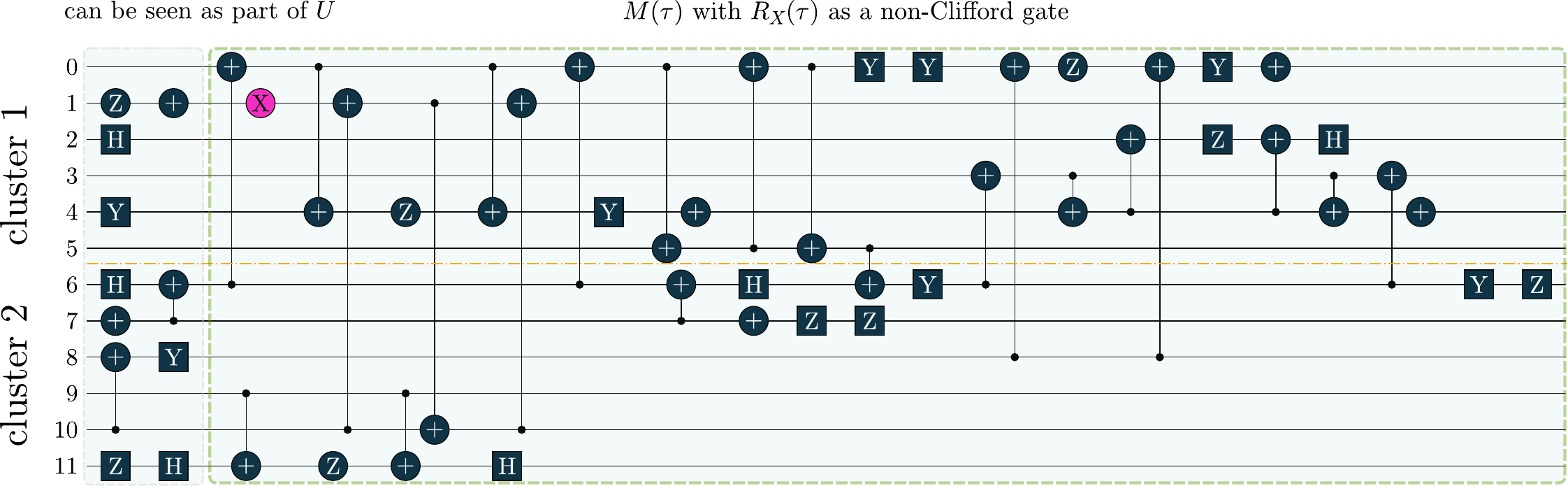}
    \caption{Optimized circuit architecture for N$_2$ at a bond distance of 2.0~{\AA}ngstrom; pink gates denote a parametrized gate and are thus non-Clifford gates. Note that gates in the beginning until the first gate with support beyond two clusters can technically be seen as an extension of the quantum circuit as they do not add correlation between clusters. One may choose to allow or not allow such gates. }
    \label{fig:n2c_200}
\end{figure*}

We further take a look at the characteristics of the learned circuits in order to check if there are some distinct circuit building blocks noticable that have been favored by the genetic algorithm used. We note that a recurring theme in the circuits discovered are $C_X$ ladder-like structures that entangle and exchange information across the qubits (\cref{fig:n2c_200}). One can also see that in the way we set up the algorithm, there is an initial part of the circuit that does not have support beyond the clusters. This means one can see this section of the virtual circuit rather as a part of $U$ than $\clascirc$, and one may choose to not allow such gates if one assumes the reference circuit $U$ is already good enough. 
An argument for leaving the gates that could be incorporated into the cluster-unitaries $U_j$ is that the algorithm has identified a part of the description of the cluster-states $\ket{\psi_j}$ that can be described classically. This is beneficial when an ansatz for the circuits $U_j$ has been established that is evaluated on real quantum computer. 
Then, thanks to $M$, the circuit for $U_j$ could be shorter.

\section{Conclusion and Outlook}\hypertarget{sec:conc-and-outlook}{}

In this work, we investigate whether a larger quantum chemical simulation can be run on smaller quantum computers without sacrificing the accuracy too much. 
This procedure is of practical interest because it allows to make use of smaller quantum computers that will be available in the near future. 
Furthermore, reducing the circuit size can allow to alleviate the problem of limited connectivity between qubits as when reduced to individual problems on less qubits, one can use better-connected parts of the architecture only.

We explored the capabilities of Clifford and near-Clifford circuits to re-entangle a molecular wavefunction that has been enforced into the form of a product state.
Within our proposed procedure, we use the energy as the only, global criterion for optimization; we deploy a genetic algorithm that mutates a set of random, initial virtual circuits using a Clifford operator pool.
\review{}{Empirical validation of our re-entangling method using Cliffords only shows that the partitioned ansatz, implemented on a smaller set of qubits, achieves similar levels of accuracy as the separable-pair ansatz.}
After successful optimization of the Clifford circuit, we ``turn on'' the parameter of \review{one}{} one the Clifford gates in the virtual circuit to further improve the energy (while continuing to adjust the parameters in the cluster circuit).
As we demonstrate, the increase in measurement cost using such near-Clifford circuits is still modest.
A worthwhile direction to follow in the future would be to perform a more concise investigation of near-Clifford operations in this context, e.g. resorting to the proposed approach in ref.~\citenum{lang2022growth} to assess the Hamiltonian growth.

Future work might include investigating more sophisticated genetic algorithms. Aside of cross-population mutations that enable broader exploration of the space of Clifford circuits, it would be interesting to include populations with permuted orbital ordering at the beginning of the optimization procedure. This is motivated by our observations in the case of BeH$_2$ in \cref{fig:beh2-perm}, where the permuted ordering is able to outperform the optimal ordering for some geometries.

Despite yielding significant improvements for H$_2$ and BeH$_2$, the improvements we see by adding the non-Clifford gate is nearly negligible in the case of N$_2$, which suggests that one non-Clifford gate is not enough in this case. Thus further work needs to be done to explore near-Clifford circuits and in particular influence of few, specific gates on the expressibility of circuits.

\section*{Acknowledgements}
We thank Abhishek Rajput for valuable comments on the manuscript.
A.A.-G. acknowledges the generous support from Google, Inc. in the form of a Google Focused Award. A.A.-G. also acknowledges support from the Canada
Industrial Research Chairs Program, the Canada 150 Research Chairs and NSERC Discovery Program. 
This work was supported by the U.S. Department of Energy under Award No. DE-SC0019374.
Resources used in preparing this research were provided, in part, by the Province of Ontario, the Government of Canada through CIFAR, Compute Canada via the SciNet HPC Consortium and companies sponsoring the Vector Institute $\langle$www.vectorinstitute.ai/partners$\rangle$.
P.S. and J.B. acknowledge support by the U.S. Department of Energy (DOE) through a quantum computing program sponsored by the Los Alamos National Laboratory (LANL) Information Science \& Technology Institute. The research at LANL was supported by the Laboratory Directed Research and Development (LDRD) program under project number 20200056DR. L.C. was partially supported by the U.S. DOE, Office of Science, Office of avanced Scientific Computing Research, under the Accelerated Research in Quantum Computing (ARQC) program.

\section*{Code and Data Availability}
Results and code for the Clifford-partitioning can be found at \url{https://github.com/philipp-q/partitioning-with-cliffords}. Furthermore code for the power method based reference methods can be found at \url{https://github.com/philipp-q/power_method_for_product_states}.

\bibliography{main.bib}

\clearpage
\onecolumngrid
\appendix
\section{Product-state enforcing reference method}\hypertarget{appendix:reference-method}{}
Given a fermionic Hamiltonian $H_\textrm{ferm}$, we use a quantum-chemical encoding~\cite{cao2019quantum,mcardle2020quantum} to obtain a qubit Hamiltonian $H_\textrm{qc}$ as a linear combination of Pauli strings. Our code for the following method is availabe at \url{https://github.com/philipp-q/power_method_for_product_states}.

Then, we use a generic wavefunction as a tensor over the qubit space, assuming a product state. For the sake of clarity, we will only show the case of two clusters and note that this can be arbitrarily extended. Then, for $\numqubits$ qubits, this gives $\ket{\psi}\in \mathbb{C}^{2^{\numqubits/2}}\otimes\mathbb{C}^{2^{\numqubits/2}}$. We use a random, initial wavefunction factorized over subsystems $A,B$ with uniformly sampled and normalized coefficients in a Pauli basis.

\begin{figure}[hb]
    \centering
    \includegraphics[width=.5\textwidth]{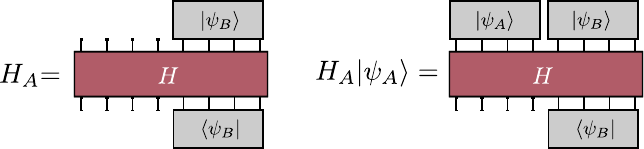}
    \caption{Hamiltonian that has been reduced over subspace $B$ and now acts only on subspace $A$.}
    \label{fig:environment}
\end{figure}

Further, we define the reduced Hamiltonians of the subsystems $A,B$ as 
\begin{align}
    H_A &= \Tr_B(H) = \braket{ (\cdot) \psi_B | H | (\cdot) \psi_B} \\
    H_B &= \Tr_A(H) = \braket{ \psi_A (\cdot) | H | \psi_A (\cdot)},
\end{align}
where $H_A$ is obtained by tracing out all subsystems except for $A$; see also \cref{fig:environment}.

To find the ground-state, we perform the following power method-type iterations using the reduced Hamiltonians for each systems, where $\ket{\psi^{(i)}} = \ket{\psi_A^{(i)}}\otimes \ket{\psi_B^{(i)}}$:
\begin{align}
    \ket{\psi_A^{(i+1)}} &= H_A \ket{\psi_A^{(i)}} - \gamma \ket{\psi_A^{(i)}}\\
    \ket{\psi_B^{(i+1)}} &= H_B \ket{\psi_B^{(i)}} - \gamma \ket{\psi_B^{(i)}}.
\end{align}

For our computations, we use $\gamma=1$ and use a convergence criterion of 
\begin{equation}
    \lvert E^{(i)} - E^{(i+1)} \rvert \le \textsc{TOL} \quad \text{where} \quad E^{(i)} = \langle\psi^{(i)}_A\psi^{(i)}_B|  H | \psi^{(i)}_A\psi^{(i)}_B  \rangle,
\end{equation}
with $\textsc{TOL}=10^{-5}$ milli-Hartree. 
We note that we keep the wavevectors as full vectors. For computationally more involved experiments than presented in this work, one could think of a representation using tensor network methods~\cite{orus2014practical}. 

\vspace{2em}



\clearpage
\section{Parameter choice for genetic algorithm}
\begin{figure}[hb]
    \centering
    \includegraphics[width=.75\linewidth]{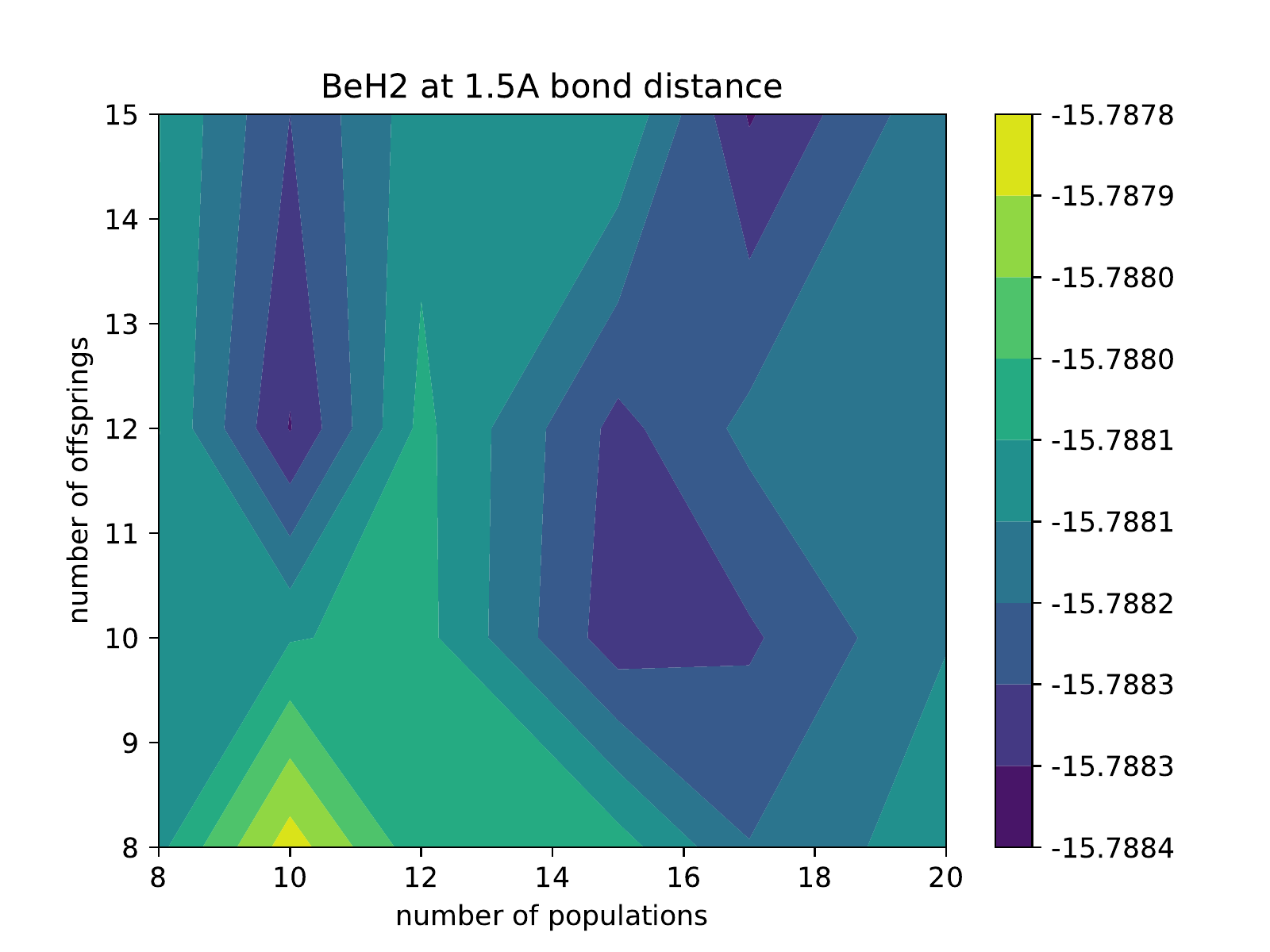}
    \caption{Parameter search for setup of genetic algorithm, exemplarity for BeH$_2$ at 1.5 {\AA}ngstrom bond distance. The color gradient depicts the obtained energy in Hartree.}
    \label{fig:parameter-search-genetic}
\end{figure}

\hypertarget{appendix:parameter-choice}{}
\review{}{
Here, we show an exemplary plot used for the hyperparameter search for the number of populations and offsprings per population for the geometric algorithm used in our procedure, as mentioned in \hyperlink{subsec:simsetup}{the simulation setup}. 
As we can see in \cref{fig:parameter-search-genetic}, the sensitivity to the parameters is around the order of $10^{-4}$ milli-Hartrees, which justifies our choice of 10--15 populations for 8--10 offsprings. 
}

\clearpage
\section{Fidelities before and after optimization}
\begin{figure}[hb]%
    \centering%
    \includegraphics[width=.85\linewidth]{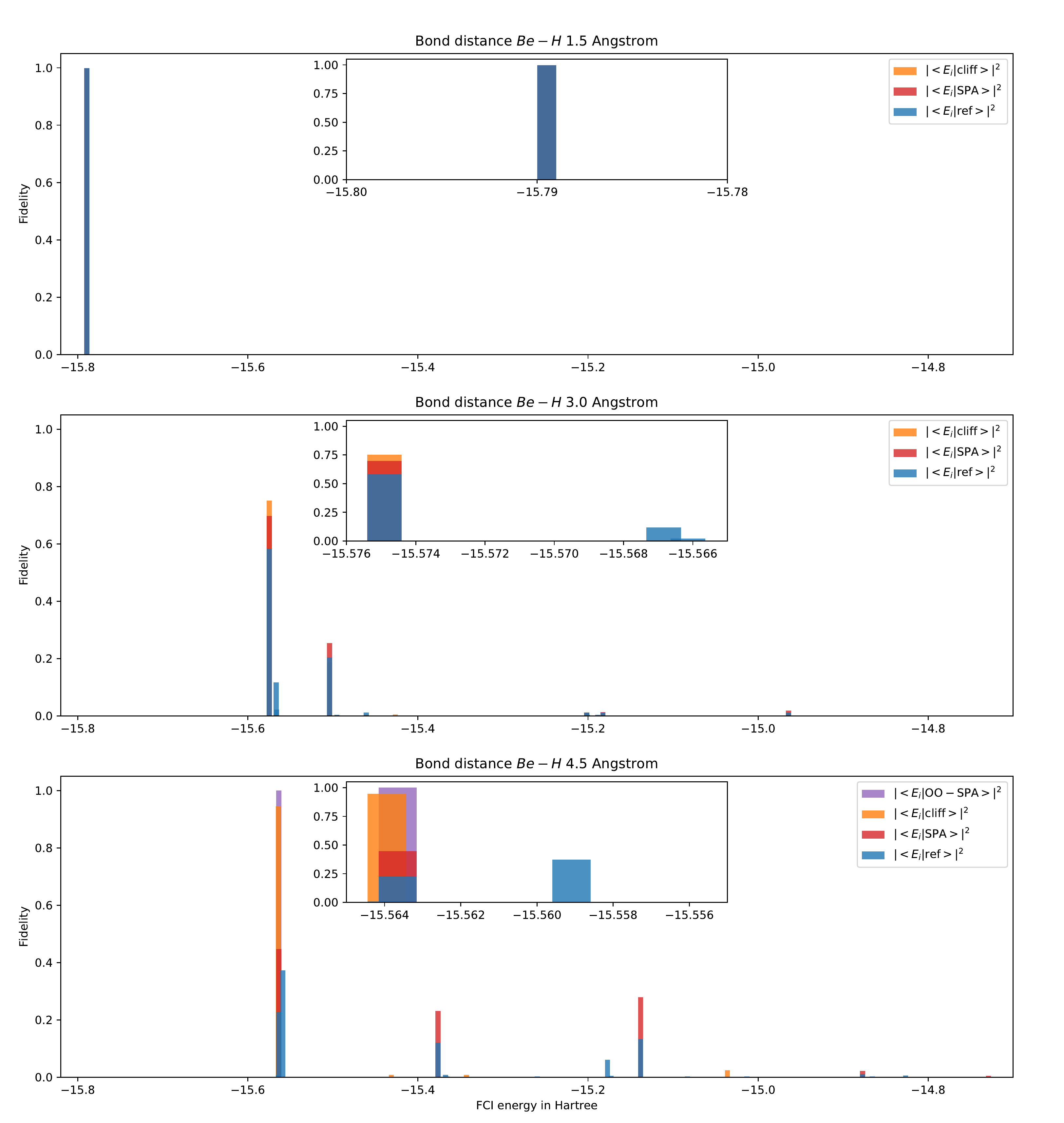}%
    \caption{Computation of fidelities between eigenstates obtained by direct diagonalization of the BeH$_2$ Hamiltonian in the PNO basis of 8~qubits in a frozen core approximation. $\mathrm{ref}\sim$ power-method based reference, $\mathrm{cliff}\sim \mathrm{ref}$ with optimized Clifford circuit. For $R=4.5$~\AA, we also provide results for orbital-optimized SPA.}%
    \label{fig:overlap-figs}%
\end{figure}

\hypertarget{appendix:fidelity-computation}{}
\review{}{Here, we provide fidelities between the FCI state obtained from exact diagonalization of the Hamiltonian with the \hyperlink{appendix:reference-method}{product-state reference method}, optimization with Clifford gates and the SPA state~\cite{kottmann2022optimized}.

We carried out the calculation exemplarily for BeH$_2$ in three geometries and show the fidelities in \cref{fig:overlap-figs} and associated energies in \cref{tab:geometries}. In the case of degenerate eigenspaces, we sum up individual overlaps for all degenerate states. 
At a bond distance of 1.5~{\AA}ngstrom, all methods show similar performance, while SPA is the most cost-effective solution thanks to the very short circuit depth in the hardcore-boson encoding~\cite{kottmann2022optimized}.
For 3.0~{\AA}ngstrom, the factorized states in ``Reference'' and SPA perform similarly. The addition of a optimized Clifford circuit led to improvements, which for this system size were small but may accumulate for larger systems.
It is not clear here whether the addition of Clifford circuits would pay off in a cost-benefit analysis.
Looking at even larger dissociation at a bond distance of 4.5~{\AA}ngstrom with a triplet ground-state and a near-degenerate singlet state, the situation is different. Since the PNOs are a bad orbital choice here (construction via MP2), standard SPA does not perform well. It is also restricted from penetrating into the triplet space (see low overlap with the ground-state in \cref{fig:overlap-figs}).
Orbital-optimization however allows to recover a lot of this error and is nearly exact. 
The ``reference'' performs better than SPA and the addition of Clifford gates allows to improve significantly upon that, however does not quite reach FCI. Both methods have overlap with inexact symmetry states and are also able to reach the triplet space. From \cref{fig:overlap-figs} (bond distance 3~\AA) we see, that while the reference has some yet significant overlap with the triplet and the subsequent singlet state, the addition of Cliffords attains high overlap with the triplet ground-state.
}

\begin{table}[hb]
\centering
\small
\begin{tabular}{c|cccc}
Bond Distance (\AA) & Reference & Reference + Cliffords & SPA & FCI  \\
\hline
1.5 & -15.7877 & -15.7881 & -15.7878 & -15.7895 \\
3.0 & -15.5318 & -15.5397 & -15.5328 & -15.5749 \\
4.5 & -15.4137 & -15.5415 & -15.3658 & -15.5636 \\
\end{tabular}
\caption{Energies in Hartree associated to overlap computation. Orbital-optimized SPA for $R=4.5$~{\AA} attains an almost exact energy of $-15.5632$~Hartree.}
\label{tab:geometries}
\end{table}

\clearpage
\section{Examples for optimized circuits}

Below, we show some exemplary circuits for each molecule that showed higher/lower performance; this data is also part of our github repository at \url{https://github.com/philipp-q/partitioning-with-cliffords}. We note that we only show the near-Clifford version with the best-performing non-Clifford gate (pink) as the respective Clifford circuit is simply given by a fixed parametrization.
We provide the optimal angles in Table~\ref{tab:optimal-angles-example-circuits}, where we notice that for nitrogen, two of the computations yield  (effectively) Clifford circuits. Beyond that, we note that in almost every case except for once, a parametrized $Z$ rotation yields the best configuration.
\begin{table}[h]
    \centering
    \begin{tabular}{c|c c c}
         & Geometry/\AA  & Angle/rad & Rotation gate\\
         \hline
        BeH$_2$ & 1.0 & -0.00466767 & $R_Z$\\
         & 3.0 &  0.00155535& $R_Z$\\
         \hline
         N$_2$ & 3.0 & -- (effectvely Clifford) & $R_Z$\\
          & 1.5 & -- (Clifford) & -- \\
          & 2.25 & 0.03207715& $R_Z$\\
         \hline
         H$_2$ & 1.4 & 2.1953$\cdot10^{-5}$& $R_X$\\
               & 2.8 & 0.00329714 & $R_Z$
    \end{tabular}
    \caption{Angles for non-Clifford gates in example circuits.}
    \label{tab:optimal-angles-example-circuits}
\end{table}

\begin{figure*}[hb]
    \centering
    \includegraphics[width=.8\linewidth]{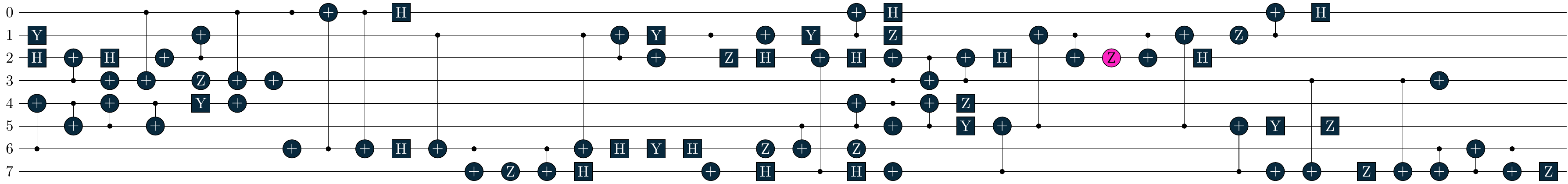}
    \caption{Better-performing near-Clifford circuit for BeH$_2$ at a bond distance of 1.0 {\AA}ngstrom.}
    \label{fig:beh2nc_10}
\end{figure*}

\begin{figure*}[hb]
    \centering
    \includegraphics[width=.8\linewidth]{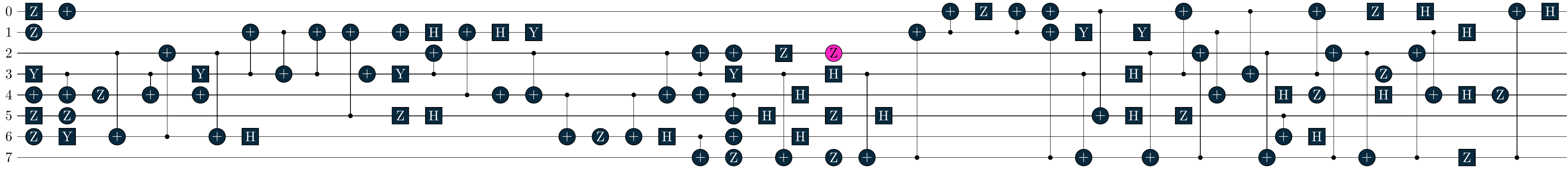}
    \caption{Near-Clifford circuit for BeH$_2$ at 3.0 {\AA}ngstrom; for BeH$_2$, the improvement is rather constant and less fluctuating than for N$_2$.}
    \label{fig:beh2nc_30}
\end{figure*}

\begin{figure*}[hb]
    \centering
    \includegraphics[width=\linewidth]{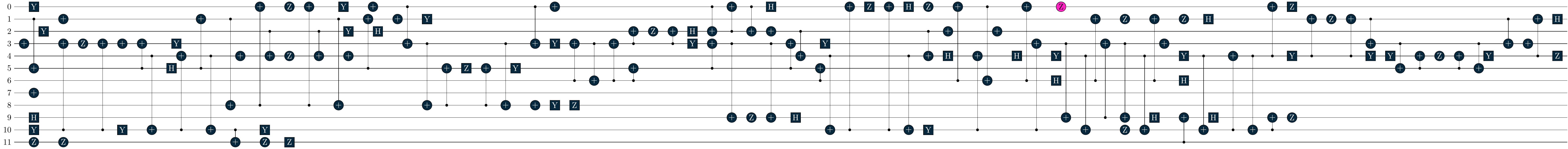}
    \caption{Near-Clifford circuit for N$_2$ at bond distance 3.0; here, substantial improvement beyond the reference and the SPA is noticed (compare to \cref{fig:n2-pes}), although the final, optimal circuit turned out to be effectively Clifford (see \cref{tab:optimal-angles-example-circuits}).}
    \label{fig:n2nc_300}
\end{figure*}

\begin{figure*}[hb]
    \centering
    \includegraphics[width=.7\linewidth]{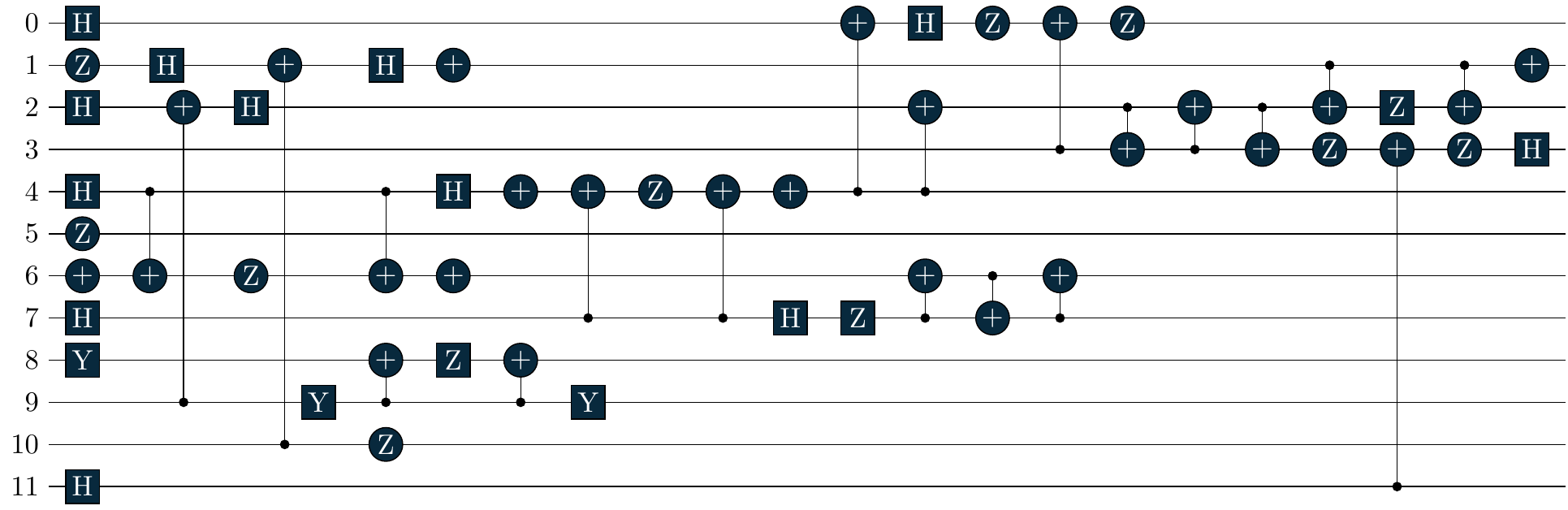}
    \caption{Poorly performing near-Clifford optimization for N$_2$ at bond distance 1.5; here, when attempting to parametrize non-Clifford gates, no improvement could have been made out, so the result remains a Clifford circuit.}
    \label{fig:n2nc_150}
\end{figure*}

\begin{figure*}[hb]
    \centering
    \includegraphics[width=.7\linewidth]{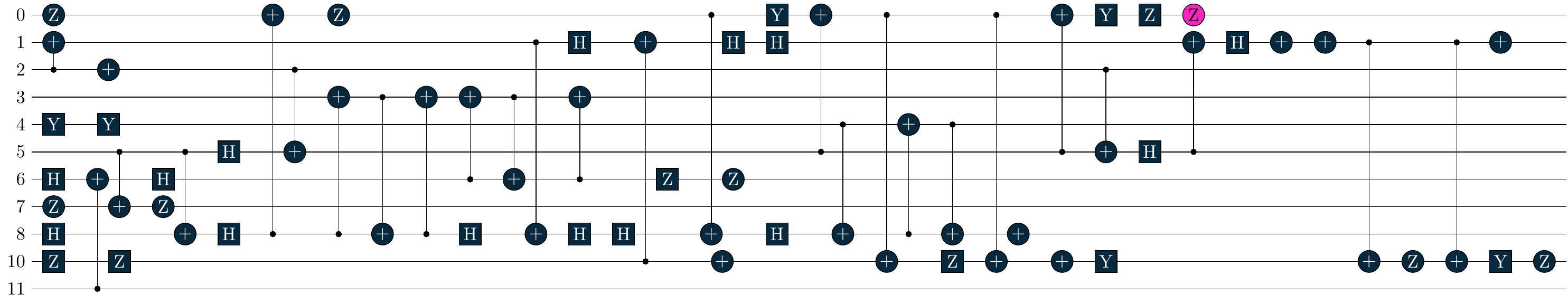}
    \caption{Well-performing near-Clifford optimization for N$_2$ at bond distance 2.25;   for this configuration, no actions have been performed on qubit 9.}
    \label{fig:n2nc_225}
\end{figure*}

\begin{figure*}[h]
    \centering
    \includegraphics[width=\linewidth]{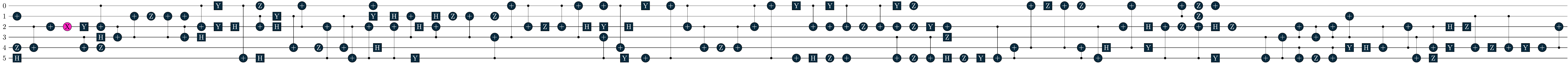}
    \caption{Near-Clifford circuit for H$_2$ at bond distance 1.4; adding and optimizing a non-Clifford gate improves error by roughly an order or magnitude here.}
    \label{fig:h2nc_140}
\end{figure*}
\begin{figure*}[h]
    \centering
    \includegraphics[width=\linewidth]{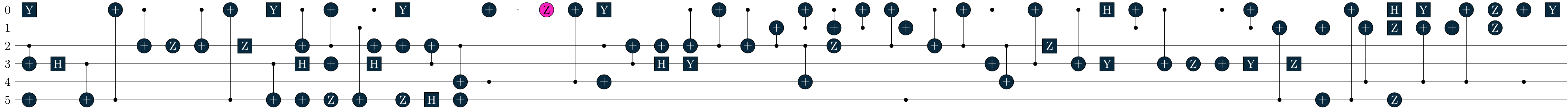}
    \caption{Near-Clifford circuit for H$_2$ at bond distance 2.8; little improvement by adding a non-Clifford gate was observed for this configuration.}
    \label{fig:h2nc_280}
\end{figure*}

\end{document}